\documentclass[aps,prfluids,reprint,onecolumn,superscriptaddress,amsmath,amssymb,longbibliography,nofootinbib]{revtex4-2}

\usepackage[T1]{fontenc}
\usepackage[utf8]{inputenc}
\usepackage[english]{babel}
\usepackage{bm}
\usepackage{graphicx}
\usepackage{booktabs}
\usepackage{xcolor}
\usepackage[colorlinks=true,linkcolor=blue!50!black,citecolor=blue!50!black,urlcolor=blue!50!black]{hyperref}

\newcommand{\bv}{\bm{u}}
\newcommand{\st}{\sigma_{\mathrm{tot}}}
\newcommand{\Pe}{\mathit{Pe}}
\newcommand{\Rey}{\mathit{Re}}
\newcommand{\St}{\mathit{St}}
\newcommand{\dd}{\,\mathrm{d}}

\begin{document}

\title{Controlling the mixing efficiency in a rotating arc-wall mixer}

\author{Teodor Burghelea}
\email{Teodor.Burghelea@univ-nantes.fr}
\affiliation{Nantes Universit\'e, CNRS, Laboratoire de Thermique et \'Energie de Nantes, LTeN, UMR 6607, F-44000 Nantes, France}

\begin{abstract}
In the rotating arc-wall mixer, the flow through a plane channel is stirred by three segments of its walls, which rotate back and forth periodically, two in phase and the third in anti-phase. Using two-dimensional simulations of a Newtonian fluid, I study how the Strouhal number $\St$ (0.25--8), the forcing ratio $\Lambda$ of the wall speed to the mean flow velocity (6.75--54), the Reynolds number $\Rey$ (0.1--20) and the P\'eclet number $\Pe$ ($10^2$ up to the grid-limited regime) set the mixing efficiency and the chaotic transport that produces it. Throughout this range the fluid crosses a chaotic saddle without regular islands, so that mixing is limited not by a lack of chaos but by the fluid that crosses the mixer without being stretched much. The contrast of the stretching field, i.e.\ how unevenly the stretching is distributed, correlates with the efficiency better than the properties of the saddle ($r=-0.90$). The efficient periods lie in a window bounded by two kinematic conditions: every fluid parcel must see at least one reversal of the walls, and the fluid at mid-channel must be displaced by about one channel width per half period. Below the window, fluid escapes in pulses through the cells of an almost steady flow; above it, fluid is held near the arcs without being folded. The window closes below $\Lambda\approx7$ and moves to faster forcing as $\Lambda$ increases. Inertia strengthens the stretching but opens a weakly stirred wake, so that the efficiency is highest at moderate inertia ($\Rey\approx8$) and decreases beyond. The efficiency decreases only slowly with $\Pe$, and a lamellar model based on the measured stretching extends the results to the P\'eclet numbers of applications. The energy cost grows as $\Lambda^2$ and does not depend on the period, which should therefore always be chosen inside the window.
\end{abstract}

\maketitle

\section{Introduction}\label{sec:intro}

Highly viscous materials, such as polymer melts and solutions, food pastes and cosmetic or pharmaceutical formulations, must often be mixed in the laminar regime, because turbulence is either out of reach or harmful to the product. Mixing then relies on laminar flows that repeatedly stretch and fold the fluid, until the striations are thin enough for molecular diffusion to homogenise them. Even simple time-periodic laminar flows can produce chaotic trajectories, and therefore efficient stretching \cite{aref1984,ottino1989,aref2002,aref2017}. The elementary mechanism is the alternation of two flows whose streamlines cross, of which linked twist maps are the minimal mathematical model \cite{wiggins2004}. In a continuous (open) flow, however, the chaos is only transient. Each fluid parcel spends a finite time near a chaotic saddle, the set of trajectories that never leave the stirred region, and is stretched by an amount that depends on how long it stays there and how close it comes to it \cite{tel2005,lt2011,jung1993}. The performance of an in-line chaotic mixer therefore depends on three properties of its flow: how strongly the saddle stretches the fluid, how long it holds it, and whether all the fluid is brought close to it. A flow that stretches strongly but lets part of the fluid through a weakly stirred channel mixes poorly, and so does a flow that holds the fluid for a long time in regions where it is not stretched.

The rotating arc-wall mixer \cite{elomari2009,PhysRevFluids.6.024502} is an in-line mixer of this kind. Three arc-shaped segments of the walls of a plane channel rotate periodically about their centres, so that the flow is stirred by moving boundaries rather than by rods or baffles placed inside it. In this respect it resembles the rotated arc mixer \cite{metcalfe2006}, a different device in which an outer cylinder rotates past arc-shaped windows cut in an inner duct, and whose Lagrangian topology has been studied for Newtonian and non-Newtonian fluids \cite{speetjens2006}. Simulations \cite{PhysRevFluids.6.024502} and experiments \cite{younes2021,youne2020} have shown that it mixes Newtonian fluids efficiently, and it has also been studied with viscoplastic \cite{moguen2023,youne2020,kiwan2025} and shear-thinning \cite{kiwan2025} fluids. Two kinematic parameters govern its operation. The Strouhal number compares the time the fluid takes to cross the mixer with the forcing period, and the forcing ratio compares the speed of the moving walls with the mean flow velocity. In an earlier study to which I contributed, El Omari et al.\ \cite{PhysRevFluids.6.024502} proposed conditions on these two parameters for the fluid to be folded by the cross-flow and to stay long enough in the mixing zone, and their simulations showed an optimal Strouhal number. Experiments \cite{kiwan2025} found the optimum at $\St\approx0.7$ for a forcing ratio of 10 and an efficiency that increases with the forcing ratio; with a shear-thinning solution, the optimum also moved to faster forcing as the forcing ratio increased.

What is still missing is a link between the control parameters and the chaotic transport that actually does the mixing. The optimum has been observed but not explained, and its dependence on the forcing ratio is unclear. The conditions of El Omari et al.\ capture the two essential ingredients, folding and retention, through estimates based on the wall speed and on the mean residence time; replacing these estimates by measured properties of the flow should sharpen the bounds and help locate the optimum. The role of inertia is unknown: experiments run at Reynolds numbers of order $10^{-1}$, whereas industrial mixers may reach $\Rey=O(10)$. The efficiency measured at a given station also depends on the P\'eclet number, which sets the scale at which diffusion takes over from stretching, and this number differs by orders of magnitude between experiments and simulations. Above all, the Lagrangian structure of the flow (the chaotic saddle, its stable and unstable manifolds, the coherent regions and the distribution of residence times) has not been characterised as a function of the control parameters. It is therefore not known through which of the three properties listed above each parameter acts.

In this paper I follow the chaotic transport of a Newtonian fluid through the rotating arc-wall mixer while varying $\St$, $\Lambda$, $\Rey$ and $\Pe$. The geometry and the numerical method are those of a companion study of shear-thinning fluids \cite{burghelea_jnnfm}. Section~\ref{sec:anatomy} describes the flow at a reference forcing: the saddle contains no regular islands, and what limits the mixing is the fluid that crosses the mixer without being stretched much. Section~\ref{sec:theory} derives two kinematic conditions that bound a window of efficient forcing periods, and Sec.~\ref{sec:StLambda} tests this window against maps of the efficiency in the $(\St,\Lambda)$ plane. Section~\ref{sec:Re} shows that the Reynolds number acts in a less direct way. Section~\ref{sec:lcs} then returns to the Lagrangian structure to explain these results: each parameter reshapes the saddle and its coherent structures differently, and a single measure, the contrast of the stretching field, correlates with the efficiency across all the runs better than the properties of the saddle. Finally, the effect of the P\'eclet number (Sec.~\ref{sec:Pe}) and the energy cost (Sec.~\ref{sec:energy}) are combined into practical rules for operating the mixer.

\section{Problem setting and methods}\label{sec:methods}

\subsection{Geometry, forcing and governing equations}\label{subsec:geometry}

\begin{figure}[!htbp]
\centering
\includegraphics[width=\textwidth]{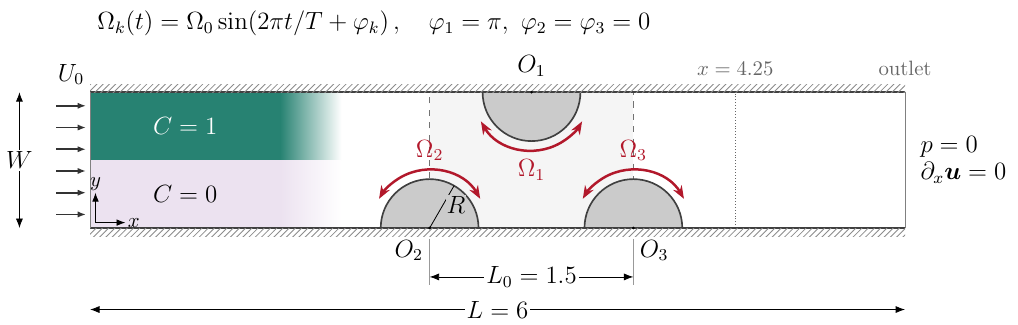}
\caption{The rotating arc-wall mixer (to scale). Two streams enter with a uniform velocity $U_0$, carrying the scalar at $C=1$ (upper half) and $C=0$ (lower half). The three arcs of radius $R$ rotate periodically about $O_{1,2,3}$, the bottom arcs in phase and the top arc in anti-phase with them; the shaded region of length $L_0$ is the active mixing zone. The dotted lines mark the stations $x=4.25$ and $x=5.5$ (outlet) where the mixing is evaluated.}
\label{fig:geometry}
\end{figure}

The mixer (Fig.~\ref{fig:geometry}) is a plane channel of width $W=1$ and length $L=6$, with $x\in[-0.5,5.5]$ and $y\in[-W/2,W/2]$. Three circular arcs of radius $R=0.36$ protrude into the channel. They are centred on the walls, at $O_1=(2.75,\,0.5)$ on the top wall and at $O_2=(2,\,-0.5)$ and $O_3=(3.5,\,-0.5)$ on the bottom wall. The active mixing zone has the length $L_0=1.5$ of the distance between $O_2$ and $O_3$. Each arc rotates rigidly about its centre with the angular velocity
\begin{equation}\label{eq:forcing}
\Omega_k(t)=\Omega_0\,\sin\!\left(\frac{2\pi t}{T}+\varphi_k\right),\qquad \varphi_2=\varphi_3=0,\quad \varphi_1=\pi ,
\end{equation}
so that the two bottom arcs rotate in phase and the top arc in anti-phase, as in the experiments \cite{younes2021,kiwan2025}. The rest of the walls is at rest. The two fluids enter at $x=-0.5$ with a uniform velocity $U_0=2$; the upper half carries a passive scalar at concentration $C=1$ and the lower half at $C=0$. At the outlet, $x=5.5$, the pressure and the normal gradients are set to zero.

The flow is incompressible and two-dimensional. The fluid is Newtonian, with viscosity $\eta$ and density $\rho=1$, and the scalar is transported by advection and diffusion:
\begin{equation}\label{eq:ns}
\nabla\cdot\bv = 0,\qquad
\rho\left(\partial_t\bv+\bv\cdot\nabla\bv\right) = -\nabla p+\eta\nabla^2\bv,\qquad
\partial_t C+\bv\cdot\nabla C = \mathcal{D}\,\nabla^2 C,
\end{equation}
with the no-slip condition on all walls, the surface of each arc moving with its rigid rotation, and no scalar flux through the walls. At $t=0$ the fluid is at rest and the two streams are stratified.

\subsection{Control parameters}\label{subsec:groups}

With $W$ and $U_0$ as reference scales and a fixed geometry, the problem depends on four dimensionless groups (Table~\ref{tab:groups}): the Strouhal number $\St=W/(U_0T)$, the forcing ratio $\Lambda=\Omega_0R/U_0$, the Reynolds number $\Rey=\rho U_0W/\eta$ and the P\'eclet number $\Pe=U_0W/\mathcal{D}$. They are varied independently through $T$, $\Omega_0$, $\eta$ and $\mathcal{D}$, respectively, with $U_0$ and $W$ fixed. $\St$ and $\Lambda$ are defined as in the experiments \cite{kiwan2025}, where $\Lambda$ is called $K$. Two derived groups are also useful. The wall Reynolds number, $\Rey_w=\Lambda\Rey$, measures the inertia of the flow driven by the arcs. The number of forcing periods during the mean transit through the mixing zone is $\bar\tau/T=\pi\St$, where $\bar\tau=V/Q=1.57$ is the mean residence time between $x=1$ and $4.75$. The Strouhal number therefore counts the reversals of the arcs that the fluid sees while it crosses the mixer.

\begin{table}[!htbp]
\centering
\caption{Control parameters. Reference values and ranges explored.}
\label{tab:groups}
\begin{tabular}{@{}llcc@{}}
\toprule
Group & Definition & Reference & Range\\
\midrule
Strouhal number & $\St=W/(U_0T)$ & 1 & 0.25--8\\
Forcing ratio & $\Lambda=\Omega_0 R/U_0$ & 27 & 6.75--54\\
Reynolds number & $\Rey=\rho U_0 W/\eta$ & 2 & 0.1--20\\
P\'eclet number & $\Pe=U_0W/\mathcal{D}$ & $2\times10^{7}$ & $10^2$--$2\times10^7$\\
Wall Reynolds number & $\Rey_w=\Lambda\Rey$ & 54 & 2.7--540\\
Arc radius & $R/W$ & 0.36 & fixed\\
\bottomrule
\end{tabular}
\end{table}

\subsection{Numerical method and validation}\label{subsec:numerics}

Equations~\eqref{eq:ns} are solved with Basilisk \cite{basilisk} on an adaptive quadtree grid, using a second-order projection method, Bell--Colella--Glaz advection \cite{bcg1989} and implicit treatment of viscous and scalar diffusion. The channel walls and the arcs are embedded boundaries \cite{johansen1998}, on which the velocity of the arc surfaces is imposed as a Dirichlet condition. The grid is adapted at every time step using a wavelet criterion on the solid fraction, the velocity and the scalar \cite{popinet2015}. A band of width 0.03 along the arcs is kept at the finest level, and the cut-cell geometry is recomputed exactly after each adaptation. The time step is limited by a CFL number of 0.8, by the wall speed ($\Delta t\le h/\Omega_0R$) and by $\Delta t\le10^{-3}$ ($10^{-4}$ for $\Omega_0\le75$). The finest cells have size $h=6/2^{10}\simeq5.9\times10^{-3}$, i.e.\ 171 cells across the channel. Each run covers $0\le t\le10$, and all quantities are computed in the time-periodic regime, over the last $\lfloor5/T\rfloor$ whole periods before $t=10$. The solver is that of the companion study of shear-thinning fluids \cite{burghelea_jnnfm}, with the viscosity and the diffusivity as parameters; at the reference parameters the two codes give bit-identical results. That study checked the approach to the periodic regime, the balance between the power injected by the arcs and the viscous dissipation, and the mean residence time, which agrees within 3.5\,\% with the value $V/Q$ imposed by mass conservation.

\begin{figure}[!htbp]
\centering
\includegraphics[width=\textwidth]{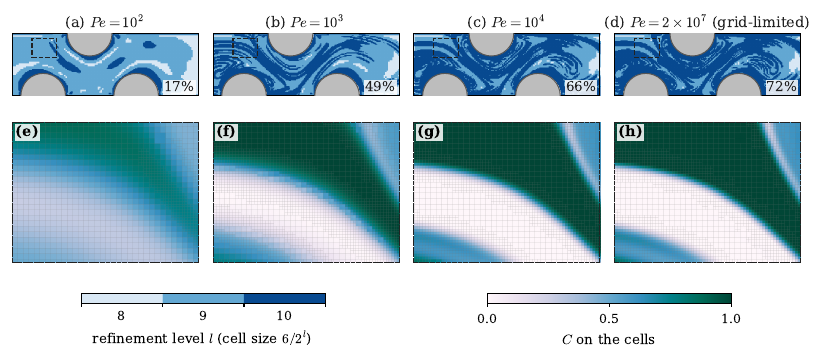}
\caption{Adaptive grid and scalar field at increasing P\'eclet number ($\St=1$, $\Lambda=27$, $\Rey=2$, level 10, $t=9$, phase 0). (a--d) Refinement level of the cells in the mixing zone $1.5\le x\le4.5$; the percentage is the fraction of its area at the finest level, and the dashed box marks the window of the lower row. (e--h) Scalar $C$ on the cells of this window, drawn without interpolation, with their edges.}
\label{fig:gridpe}
\end{figure}

At the reference parameters, the penetration length of the forcing (Sec.~\ref{subsec:diagnostics}) changes by only 0.2\,\% between grid levels 9 and 10, whereas the outlet mixing efficiency is $M=0.700$, 0.654 and 0.622 at levels 9, 10 and 11. The velocity field is thus converged, but the absolute value of $M$ is not: at $\Pe=2\times10^7$ the striations are homogenised when they reach the grid scale. The viscous dissipation, from which the power is computed (Sec.~\ref{sec:energy}), converges more slowly than the velocity: at the reference parameters it differs by 15\,\% between levels 9 and 10 and by 7\,\% between levels 10 and 11, because the stress is singular at the feet of the arcs \cite{burghelea_jnnfm}. The power coefficients of Sec.~\ref{sec:energy} are therefore uncertain by about 10\,\% in absolute value; their variations with $\St$, $\Lambda$ and $\Rey$, compared at the same resolution, are much smaller than this. The run at $\Rey=20$, where the viscous layer on the arcs, $(\nu/\Omega_0)^{1/2}\simeq0.026$, spans only four cells at level 10, was repeated at levels 9 and 11. Between levels 10 and 11, the penetration length changes by 0.2\,\%, the mean residence time by 1.4\,\% and the mean stretching rate by 5\,\%, and the fraction of the flux stretched by less than $e^2$ does not change (0.148). The efficiency decreases with the resolution as at the reference parameters, from 0.691 to 0.622 and 0.586 at levels 9, 10 and 11; its ratio to the reference value, 0.99, 0.95 and 0.94, does not move towards one, so that the loss of efficiency at $\Rey=20$ (Sec.~\ref{sec:Re}) is not an effect of the resolution. The dissipation differs by 21\,\% between levels 9 and 10 and by 4.5\,\% between levels 10 and 11, and its increase with respect to $\Rey=2$, 14, 17 and 20\,\% at levels 9, 10 and 11, is robust. The gain at $\Rey=8$ was computed at level 10 only, and, being of the same order as the change of this ratio with the grid, is indicative. Stretching at a rate $\lambda$ thins the striations down to the Batchelor scale $\ell_B=(\mathcal{D}/\lambda)^{1/2}$ \cite{batchelor1959}, which the grid resolves if $\ell_B\gtrsim2h$, i.e.\ if $\Pe\lesssim U_0W/(4h^2\lambda)$. With the measured stretching rate, about 4, this gives $\Pe\lesssim3\times10^3$ at level 10. Figure~\ref{fig:gridpe} shows how the grid responds. The wavelet criterion refines wherever the scalar varies on the scale of the cells, so that the finest level covers 17, 49 and 66\,\% of the mixing zone at $\Pe=10^2$, $10^3$ and $10^4$, and 72\,\% at $\Pe=2\times10^7$. At $\Pe\le10^3$, the fronts between the two streams extend over several of the finest cells (Fig.~\ref{fig:gridpe}e,f); from $\Pe=10^4$ they are about two cells wide, and the fields at $10^4$ and $2\times10^7$ are almost indistinguishable (Fig.~\ref{fig:gridpe}g,h). At larger P\'eclet numbers, numerical diffusion at the grid scale replaces molecular diffusion, and the results depend on the grid but no longer on $\Pe$. The reference value $\Pe=2\times10^7$ represents this grid-limited regime, which stands in for the very large P\'eclet numbers of the experiments ($\Pe\sim10^9$ \cite{kiwan2025}). All comparisons in $\St$, $\Lambda$ and $\Rey$ are made in this regime, at a fixed resolution. The runs at $\Pe=10^2$--$10^4$ show how the conclusions depend on $\Pe$, and they give the effective P\'eclet number of the grid, about $2\times10^4$ at level 10 (Sec.~\ref{sec:Pe}). The Lagrangian diagnostics described below use only the velocity field and are not affected by this limitation.

\subsection{Diagnostics}\label{subsec:diagnostics}

\paragraph{Mixing efficiency} The mixture delivered by the mixer is characterised at the outlet, $x=5.5$, by the flux-weighted standard deviation of the concentration over the channel width and over the periodic regime,
\begin{equation}\label{eq:sigmatot}
\st=\left[\frac{\iint u\,(C-\bar C)^2\dd y\dd t}{\iint u\dd y\dd t}\right]^{1/2},\qquad
\bar C=\frac{\iint u\,C\dd y\dd t}{\iint u\dd y\dd t},
\end{equation}
i.e.\ by the variance of the fluid collected at the outlet. The mixing efficiency is $M=1-2\st$: it is 0 for fully segregated streams and 1 for perfect mixing. Without diffusion, $\st$ is the same at every cross-section of a time-periodic flow, so it decreases downstream only through diffusion. With $q(t)=\int u\dd y$ and the composition delivered at time $t$, $m(t)=\int uC\dd y/q$, it splits exactly into a spatial and a temporal part \cite{burghelea_jnnfm},
\begin{equation}\label{eq:split}
\st^2=\frac{\int q\,\sigma_q^2\dd t}{\int q\dd t}+\frac{\int q\,(m-\bar C)^2\dd t}{\int q\dd t},\qquad \sigma_q^2=\frac{\int u\,(C-m)^2\dd y}{q},
\end{equation}
the first measuring the non-uniformity of the delivered fluid across the section and the second the fluctuation of its composition in time. The penetration of the wall forcing into the bulk is measured by the e-folding length $\ell$ of a least-squares exponential fit of the oscillating velocity $u'_{\mathrm{rms}}=[\overline{(u-\bar u)^2}+\overline{(v-\bar v)^2}]^{1/2}$, binned as a function of the distance $d$ to the nearest arc, over $0.01\le d\le0.1$ \cite{burghelea_jnnfm}. The energy cost is measured by the viscous dissipation over the active zone $1.5\le x\le4.25$, averaged over a period, Eq.~\eqref{eq:Pdef} of Sec.~\ref{sec:energy}; in the periodic regime it accounts for 92--96\,\% of the power injected by the arcs \cite{burghelea_jnnfm}, the rest being dissipated outside this zone.

\paragraph{Lagrangian transport} All Lagrangian quantities are computed from the phase-averaged periodic velocity field (100 phases per period, averaged over the periods within $5\le t<10$). Trajectories are integrated with a fourth-order Runge--Kutta scheme and a step of about $5\times10^{-4}$, the same in time units for every period. The methods are those of the companion study \cite{burghelea_jnnfm}, except that all time horizons are fixed in time units rather than in numbers of periods, so that runs with different periods can be compared. Three groups of quantities are used.

\emph{Transit of the fluid.} Tracers released at the entrance of the mixing zone, $x=1$, at all phases and weighted by the local flux, are followed until they reach $x=4.75$. They give the flux-weighted residence-time distribution $E(\tau)$ and its cumulative distribution $F(\tau)$; the shortest residence time $\tau_{\min}$, defined by $F=10^{-3}$; the coefficient of variation $c_v$ of $\tau$; and the log-stretch $s$ of a material element carried by each parcel through the mixer, with its median $\tilde s$ and the fraction of the flux that is poorly stretched ($s<2$). The mean stretching rate of the fluid crossing the mixer is the flux-weighted mean $\bar\lambda=\langle s/\tau\rangle$, and $\langle\tau\rangle$ denotes the flux-weighted mean residence time.

\emph{Chaotic saddle.} In an open flow, the trajectories that stay longest accumulate on a chaotic saddle and along its unstable manifold \cite{jung1993,lt2011}. To locate them, a uniform grid of $8.7\times10^4$ tracers (spacing $0.006\simeq h$) is seeded over the fluid between $x=1$ and $4.75$ at phase 0 and advected for eight time units; each tracer is removed when it crosses $x=4.75$. The positions of the remaining tracers (the survivors) at $t=kT$ form the stroboscopic, or Poincar\'e, sections of the saddle. Such sections show where the long-lived fluid goes, and which regions it cannot reach, but not how strongly it is stretched; stretching statistics predict the spreading of a tracer better \cite{swanson1990}, and both are used here. While the survivors are in the bulk, the survival probability $P_s(t)$ of the tracers decays as $e^{-\kappa t}$, which defines the escape rate $\kappa$ \cite{kantz1985}. It is fitted between $t=T$ and the time $t_w$ at which half of the survivors lie within $0.02$ of a wall. The Lyapunov exponent of the saddle, $\lambda_s$, is the mean growth rate of the log-stretch of the survivors that are away from the walls, over the same interval. The Kantz--Grassberger relation then gives the dimension of the unstable manifold, $D_u=2-\kappa/\lambda_s$ \cite{kantz1985,lt2011}. It lies between 1, for a smooth curve (fast escape or weak stretching), and 2, for a set that fills the plane. These rates depend somewhat on the fit window: over the fixed window $0.5\le t\le3$ used in the companion study \cite{burghelea_jnnfm}, the reference run gives $\kappa=1.30$ and $\lambda_s=4.8$ instead of 1.25 and 4.6, with the same $D_u=1.73$.

\emph{Coherent structures.} The forward finite-time Lyapunov exponent (FTLE) of a fluid element at $\bm x_0$ is
\begin{equation}\label{eq:ftle}
\lambda_F(\bm x_0)=\frac{1}{t_F}\ln\sqrt{\mu_{\max}\big(\mathsf{C}(\bm x_0)\big)},\qquad \mathsf{C}=(\nabla\bm\phi)^{\mathsf T}\,\nabla\bm\phi,
\end{equation}
where $\bm\phi$ is the flow map from phase 0 over $t_F=1$ time unit, $\mathsf{C}$ the Cauchy--Green tensor and $\mu_{\max}$ its largest eigenvalue; the backward FTLE is obtained in the same way with the flow map over $-t_F$. Both are computed on a grid of spacing $2h$ covering $1.5\le x\le4.5$, the gradient of the flow map by finite differences. Their ridges are the hyperbolic Lagrangian coherent structures: finite-time approximations of the stable and unstable manifolds of the saddle \cite{slm2005,haller2015}, which divide the flow into lobes that are transported as units \cite{branicki2010}. Elliptic coherent regions, i.e.\ islands where the fluid rotates without being stretched, are identified as connected regions of at least 25 grid cells in which both FTLE are below half their mean; $A_{\mathrm{ell}}$ denotes their area fraction. The fluid that will not be stretched during the next time unit is measured by $A_w$, the area fraction where the forward FTLE is below half its mean, over the whole zone or only downstream of the last arc ($x>3.9$). The contrast of the stretching field is
\begin{equation}\label{eq:H}
H=\frac{P_{95}(\lambda_F)}{P_{50}(\lambda_F)},
\end{equation}
where $P_q$ denotes the $q$th percentile of the forward FTLE over the fluid in $1.5\le x\le4.5$. It is close to 1 when the stretching is uniform, and large when the stretching is concentrated on a few ridges separated by weakly stretched regions. Finite-time exponents fluctuate widely about their long-time value, so that their distribution, not only its mean, characterises the stretching \cite{johnson2015}; $H$ is the simplest measure of its spread. Finally, the density of folds is measured by the number of FTLE ridges that cross the channel, averaged over vertical lines between the arcs.

\subsection{Simulations}\label{subsec:runs}
Table~\ref{tab:runs} lists the simulations. The Strouhal number is varied at six forcing ratios, $\Lambda=9.54$, 13.5, 19.08, 27, 38.16 and 54 ($\Omega_0=53$, 75, 106, 150, 212 and 300, in steps of about $\sqrt2$), and at $\Lambda=6.75$, below the critical forcing ratio predicted in Sec.~\ref{sec:theory}. The Reynolds and P\'eclet numbers are varied at the reference forcing, $\St=1$ and $\Lambda=27$. The Lagrangian structure is computed for the three one-parameter series that pass through the reference point: $\St$ at $\Lambda=27$, $\Rey$ at $\St=1$, and $\Lambda$ at $\St=1$ (18 runs in total). Ten of the runs, the Strouhal series at $\Lambda=27$ and the runs at $\St=1$ with $\Lambda=13.5$ and 54, as well as the grid checks at the reference parameters, are the Newtonian runs of the companion study \cite{burghelea_jnnfm}, where their outlet efficiency, penetration length and power serve as the reference for the shear-thinning fluids. All the other runs, and the analyses of the chaotic saddle and its coherent structures, of the window of efficient periods and of the effects of the Reynolds and P\'eclet numbers, are new.

\begin{table}[!htbp]
\centering
\caption{Simulations. Unless stated, $\Rey=2$ and $\Pe=2\times10^7$.}
\label{tab:runs}
\begin{tabular}{@{}lll@{}}
\toprule
Series & Fixed & Varied\\
\midrule
$\St$ at $\Lambda=27$ & $\Lambda=27$ & $\St=0.25$, 0.5, 0.67, 0.8, 1, 1.6, 2, 4\\
$\St$ at $\Lambda=9.54$, 13.5, 19.08, 38.16 & each $\Lambda$ & $\St=0.25$, 0.5, 1, 2, 4\\
$\St$ at $\Lambda=54$ & $\Lambda=54$ & $\St=0.25$, 0.5, 1, 2, 4, 6, 8\\
$\St$ at $\Lambda=6.75$ & $\Lambda=6.75$ & $\St=0.25$, 0.5, 1, 2\\
$\Lambda$ & $\St=1$ & $\Lambda=6.75$, 9.54, 13.5, 19.08, 27, 38.16, 54\\
$\Rey$ & $\St=1$, $\Lambda=27$ & $\Rey=0.1$, 0.5, 2, 8, 20\\
$\Pe$ & $\St=1$, $\Lambda=27$ & $\Pe=10^2$, $10^3$, $3\times10^3$, $10^4$, $2\times10^7$\\
\bottomrule
\end{tabular}
\end{table}

\section{Anatomy of the chaotic flow at the reference forcing}\label{sec:anatomy}

At the reference parameters ($\St=1$, $\Lambda=27$, $\Rey=2$), the flow becomes periodic after a few forcing periods. Over $5\le t<10$, the mixing efficiency computed period by period varies by only about $10^{-3}$, and its value at the outlet is $M=0.654$. The arcs drive a cellular cross-flow whose direction alternates with their rotation. The oscillating velocity decays away from the arcs over a penetration length $\ell=0.064W$, but the cross-flow it drives reaches the middle of the channel (Sec.~\ref{subsec:linear}). The fluid crosses the mixing zone in a mean time $\langle\tau\rangle=1.53$, about three forcing periods, and is stretched at a mean rate $\bar\lambda=4.34$.

\begin{figure}[!htbp]
\centering
\includegraphics[width=\textwidth]{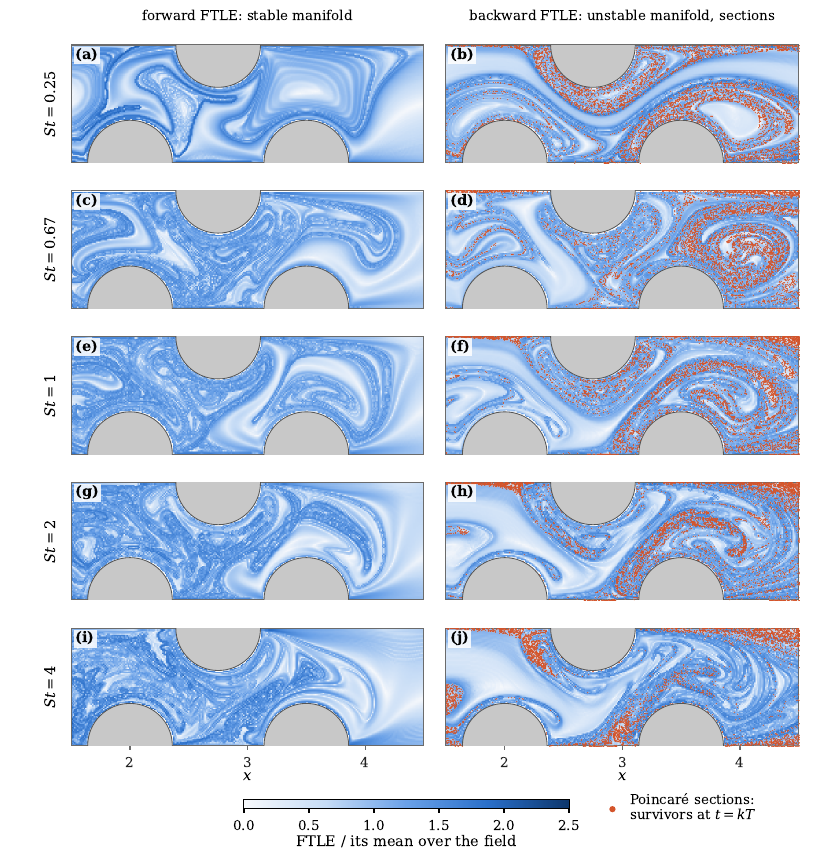}
\caption{Chaotic saddle and coherent structures at $\Lambda=27$, $\Rey=2$, for $\St=0.25$, 0.67 (optimum), 1 (reference), 2 and 4 (rows). Left: forward FTLE over one time unit from phase 0, relative to its mean over the field; its ridges are the stable manifold of the saddle. Right: backward FTLE (ridges: unstable manifold), with the stroboscopic sections of the saddle superposed (dots: tracers still in the mixing zone at $t=kT$, $1.5\le kT\le3$). Grey: arcs.}
\label{fig:manst}
\end{figure}

\begin{figure}[!htbp]
\centering
\includegraphics[width=\textwidth]{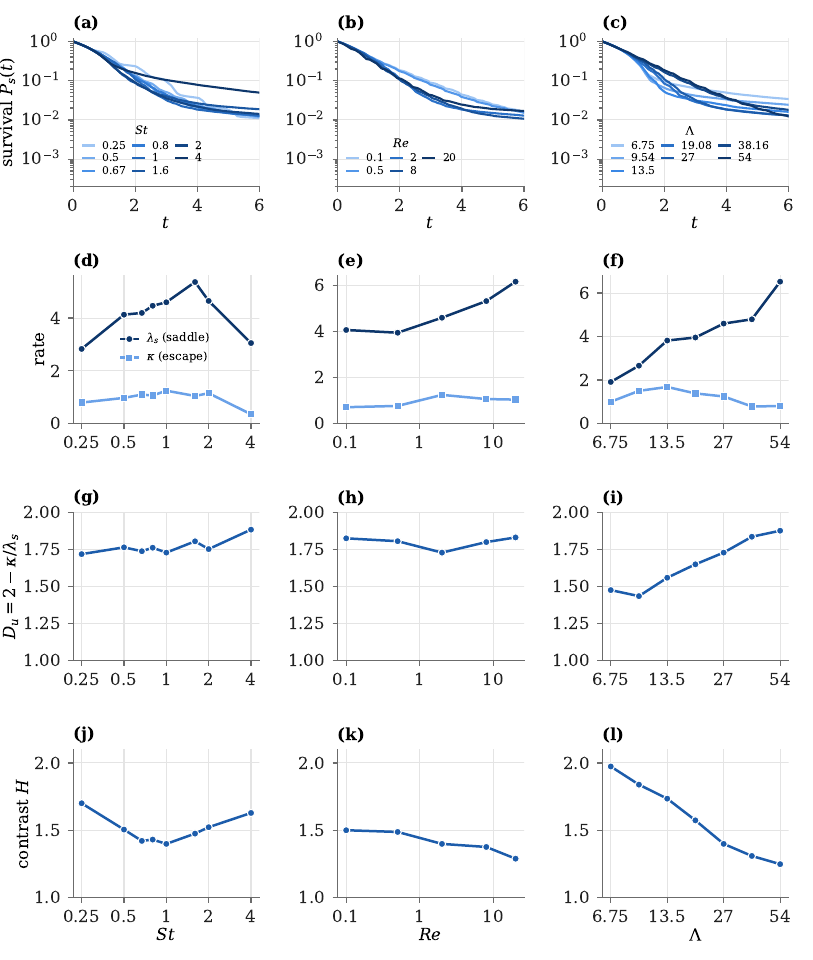}
\caption{Chaotic saddle versus the Strouhal number ($\Lambda=27$, left), the Reynolds number ($\St=1$, $\Lambda=27$, centre) and the forcing ratio ($\St=1$, right), $\Rey=2$ unless varied. (a--c) Survival probability of tracers seeded uniformly in the mixing zone; (d--f) escape rate $\kappa$ and Lyapunov exponent of the saddle $\lambda_s$; (g--i) dimension of the unstable manifold $D_u=2-\kappa/\lambda_s$; (j--l) contrast of the stretching field, $H$, the ratio of the 95th percentile to the median of the forward FTLE.}
\label{fig:chaosstats}
\end{figure}

The stretching is organised by the manifolds of a chaotic saddle (Fig.~\ref{fig:manst}e,f). The ridges of the backward FTLE trace the unstable manifold, a set of folded filaments that wind around the downstream bottom arc and through the gap between this arc and the top arc. The ridges of the forward FTLE trace the stable manifold, along which incoming fluid is drawn towards the saddle. Three to four ridges of each kind cross the channel at any section between the arcs, and the fluid between two neighbouring ridges is carried as a lobe from one fold to the next. As expected, the Poincar\'e sections of the saddle lie on the ridges of the backward FTLE. The survival probability (Fig.~\ref{fig:chaosstats}a) decays exponentially with an escape rate $\kappa=1.25$: 42\,\% of the tracers remain after one time unit, 11\,\% after two and 1.8\,\% after four. Meanwhile, the survivors are stretched at the Lyapunov exponent of the saddle, $\lambda_s=4.6$. The unstable manifold therefore has dimension $D_u=1.73$: the saddle is a dense set that nearly fills the plane, and the flow is in the regime of strong transient chaos.

The flow contains no regular islands of significant size. The regions where neither FTLE exceeds half its mean, which would contain such islands, cover only 0.4\,\% of the mixing zone, in two small patches. This threshold is a practical criterion rather than a rigorous detection of elliptic coherent structures \cite{haller2015,haller2013}: it would miss islands smaller than the minimum size retained, 25 grid cells, i.e.\ a radius of about $0.017W$. By reversing every half period, the forcing breaks up the closed streamlines of the instantaneous flow, as intended in the design of the mixer \cite{PhysRevFluids.6.024502}. The coherent structures that matter are therefore not islands, but the hyperbolic structures themselves and the weakly stretched regions between them. The forward FTLE is below half its mean over 10\,\% of the mixing zone, and over 15\,\% of the region downstream of the last arc; the contrast of the stretching field is $H=1.40$. After about seven periods ($t_w=3.5$), most survivors lie within $0.02$ of a wall: 71\,\% of them after four time units, and 96\,\% after eight. The long-lived fluid that departs from the exponential decay is thus fluid held near the no-slip walls, where it is only weakly stretched \cite{gouillart2007,gouillart2008}.

\begin{figure}[!htbp]
\centering
\includegraphics[width=\textwidth]{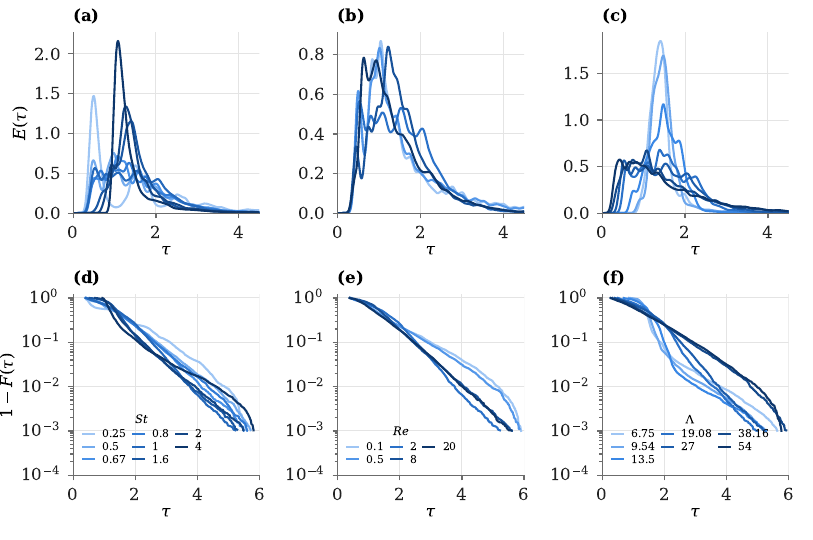}
\caption{Flux-weighted residence-time distribution between $x=1$ and $4.75$ for the same three series as Fig.~\ref{fig:chaosstats}: (a--c) density $E(\tau)$ (kernel estimate); (d--f) fraction of the flux still in the mixing zone, $1-F(\tau)$.}
\label{fig:rtd}
\end{figure}

The residence-time distribution (Fig.~\ref{fig:rtd}) describes the same dynamics from the point of view of the fluid that crosses the mixer. The fastest fluid takes $\tau_{\min}=0.40$, a quarter of the mean time. The distribution is broad ($c_v=0.51$) and has an exponential tail, the signature of a hyperbolic saddle; 3.8\,\% of the flux stays longer than twice the mean time. Accordingly, the log-stretch of the fluid crossing the mixer is broadly distributed (quartiles 3.2, 5.6 and 8.8), and 15\,\% of the flux leaves with $s<2$, i.e.\ stretched by less than a factor $e^2$. This poorly stretched fluid has either crossed quickly or travelled along the weakly stretched regions, and it is what limits the mixing.

The same quantities are given for all the runs analysed in Table~\ref{tab:lagr}, and they are discussed in Sec.~\ref{sec:lcs}.

\begin{table}[!htbp]
\centering
\caption{Lagrangian characterisation of the three series through the reference point ($\St=1$, $\Lambda=27$, $\Rey=2$, listed once). $M$: outlet mixing efficiency; $\tau_{\min}$, $c_v$: shortest residence time and coefficient of variation of the residence time; $\kappa$, $\lambda_s$, $D_u$: escape rate, Lyapunov exponent and dimension of the unstable manifold of the saddle; $H$: contrast of the forward FTLE; $A_w$: area fraction where the forward FTLE is below half its mean; $f_{s<2}$, $\tilde s$: fraction of the flux with log-stretch $s<2$, and median log-stretch, of the fluid crossing the mixer.}
\label{tab:lagr}
\begin{tabular}{@{}lccccccccccc@{}}
\toprule
 & $M$ & $\tau_{\min}$ & $c_v$ & $\kappa$ & $\lambda_s$ & $D_u$ & $H$ & $A_w$ & $f_{s<2}$ & $\tilde s$\\
\midrule
$\St=0.25$ & 0.480 & 0.38 & 0.73 & 0.80 & 2.83 & 1.72 & 1.70 & 0.097 & 0.44 & 2.8\\
$\St=0.5$  & 0.632 & 0.41 & 0.59 & 0.97 & 4.13 & 1.76 & 1.51 & 0.100 & 0.22 & 4.5\\
$\St=0.67$ & 0.695 & 0.43 & 0.53 & 1.10 & 4.19 & 1.74 & 1.42 & 0.101 & 0.18 & 5.0\\
$\St=0.8$  & 0.687 & 0.41 & 0.55 & 1.06 & 4.47 & 1.76 & 1.43 & 0.098 & 0.18 & 5.0\\
reference  & 0.654 & 0.40 & 0.51 & 1.25 & 4.60 & 1.73 & 1.40 & 0.100 & 0.15 & 5.6\\
$\St=1.6$  & 0.626 & 0.53 & 0.39 & 1.05 & 5.37 & 1.81 & 1.48 & 0.099 & 0.13 & 4.9\\
$\St=2$    & 0.609 & 0.72 & 0.40 & 1.15 & 4.65 & 1.75 & 1.52 & 0.109 & 0.15 & 4.8\\
$\St=4$    & 0.329 & 0.94 & 0.47 & 0.35 & 3.05 & 1.88 & 1.63 & 0.126 & 0.25 & 3.4\\
\midrule
$\Rey=0.1$ & 0.640 & 0.43 & 0.66 & 0.71 & 4.07 & 1.83 & 1.50 & 0.110 & 0.24 & 4.5\\
$\Rey=0.5$ & 0.641 & 0.42 & 0.64 & 0.77 & 3.95 & 1.81 & 1.49 & 0.105 & 0.22 & 4.6\\
$\Rey=8$   & 0.699 & 0.39 & 0.49 & 1.07 & 5.32 & 1.80 & 1.38 & 0.112 & 0.10 & 5.7\\
$\Rey=20$  & 0.622 & 0.38 & 0.57 & 1.04 & 6.16 & 1.83 & 1.29 & 0.128 & 0.15 & 5.7\\
\midrule
$\Lambda=6.75$  & 0.201 & 0.90 & 0.32 & 1.00 & 1.91 & 1.48 & 1.97 & 0.191 & 0.47 & 2.1\\
$\Lambda=9.54$  & 0.347 & 0.86 & 0.28 & 1.50 & 2.66 & 1.44 & 1.84 & 0.158 & 0.36 & 2.7\\
$\Lambda=13.5$  & 0.534 & 0.69 & 0.30 & 1.68 & 3.82 & 1.56 & 1.73 & 0.124 & 0.23 & 3.5\\
$\Lambda=19.08$ & 0.576 & 0.50 & 0.44 & 1.39 & 3.96 & 1.65 & 1.57 & 0.116 & 0.20 & 4.4\\
$\Lambda=38.16$ & 0.702 & 0.32 & 0.63 & 0.79 & 4.79 & 1.84 & 1.31 & 0.078 & 0.14 & 6.4\\
$\Lambda=54$    & 0.830 & 0.26 & 0.69 & 0.81 & 6.53 & 1.88 & 1.25 & 0.057 & 0.12 & 8.1\\
\bottomrule
\end{tabular}
\end{table}

The reference flow thus defines the problem. The saddle stretches strongly, but the mixing at the outlet is set by the fluid that escapes it: fluid that crosses before the arcs have folded it, fluid carried along the weakly stretched regions, and fluid held near the walls. The control parameters affect the mixing by changing how much fluid falls into each of these categories. The next section uses the kinematics of the cross-flow to derive two conditions that keep the first category empty.

\section{Kinematic criteria for mixing}\label{sec:theory}

\subsection{The conditions of El Omari et al.}\label{subsec:elomari}
In our earlier study, El Omari et al.\ \cite{PhysRevFluids.6.024502} defined the forcing ratio exactly as $\Lambda$ is defined here (their $K_0=R\Omega_0/v_{\mathrm{bulk}}$, with $R$ the radius of the cylinders) and derived two conditions. Their \emph{folding condition} requires that a tracer next to a rotating arc be displaced across at least one channel width during a half period, at the mean wall speed $(2/\pi)\Omega_0R$:
\begin{equation}\label{eq:EOfold}
\Lambda\ge\pi\St .
\end{equation}
Their \emph{mixing condition} requires the residence time in a mixing zone of length $\zeta W$ to exceed the time needed for diffusion across striations of thickness $W/2^n$, after $n$ folds. This requires $n\ge\ln(\Pe/\zeta)/\ln4$ folds, while the number of folds is bounded by the transverse displacement during the residence time, $n\le\zeta\bar\Omega R/U_0$, where $\bar\Omega$ is the mean of $|\Omega(t)|$ over the residence time. The result is (their Eq.~22)
\begin{equation}\label{eq:EOmix}
\Lambda\ge\frac{\pi\St}{\ln4}\,\frac{\ln(\Pe/\zeta)}{2\zeta\St+\sin^2(\pi\zeta\St)}
\;\xrightarrow{\;\zeta\St\gg1\;}\;\frac{\pi\ln(\Pe/\zeta)}{2\zeta\ln4},
\end{equation}
i.e.\ a minimum forcing ratio of about 7.4 for their values $\zeta=2$ and $\Pe=10^6$, nearly independent of $\St$. With $N$ whole periods in the residence time $\zeta W/U_0$ and a remainder of at most half a period, the mean of $|\sin(2\pi t/T)|$ is $[2N+\sin^2(\pi\zeta\St)]/(\pi\zeta\St)$ (their Eq.~20); Eq.~\eqref{eq:EOmix}, the curve drawn in their maps, is the continuous form of this estimate, with $N=\zeta\St$; keeping $N=\lfloor\zeta\St\rfloor$ makes the bound grow as $\St\to0$, as $\ln(\Pe/\zeta)/(\pi\ln4\,\zeta^2\St)$. The two forms coincide for $\zeta\St\gtrsim1$ and have the same large-$\St$ limit, which is the one used below. Their simulations, 90 runs at $\Pe=10^6$, showed good mixing in the region bounded by Eqs.~\eqref{eq:EOfold} and \eqref{eq:EOmix}, with an optimal Strouhal number inside it.

These conditions identify the two ingredients on which the analysis below is built, folding and retention. Since the structure of the flow inside the mixer was not known when they were derived, they rely on three estimates, which the simulations of Sec.~\ref{sec:anatomy} now allow me to replace by measured quantities. First, the velocity of the cross-flow is estimated by the wall speed; the fluid that must be folded, initially at mid-channel, actually moves more slowly, by a factor $\alpha$ measured below. Second, the folds are counted from the accumulated transverse displacement, so that their number grows with $\Lambda$; in a slow, nearly reversible flow, most of the displacement during one half period is undone during the next, and a fold is made each time the arcs reverse. Third, a single residence time is assigned to all the fluid; the outlet variance is controlled by the fluid that crosses fastest, which the residence-time distribution gives. The criteria derived below keep the structure of the conditions of El Omari et al.\ and replace these three estimates by measured properties of the flow.

\subsection{Linear, quasi-steady response of the cross-flow}\label{subsec:linear}
The transverse velocity on the centreline, $y=0$, oscillates at the forcing frequency. The amplitude of its first harmonic, $v_1(x)$, is largest at $x=3.13$, in the gap between the top arc and the downstream bottom arc. Scaled by the wall speed, this maximum is
\begin{equation}\label{eq:alpha}
\alpha=\max_x v_1/(\Omega_0R),
\end{equation}
and its value is 0.283, 0.284 and 0.282 for $T=0.5$, 1 and 0.25 at $\Omega_0=150$, and 0.291 at $\Omega_0=75$. The cross-flow in the bulk is thus proportional to the wall speed and independent of the period, even though the wall Reynolds number is 54. In other words, the bulk responds linearly and quasi-steadily to the forcing, and the velocity is approximately the superposition $\bv\simeq U_0\bm{u}_b(\bm{x})+\Omega(t)R\,\bm{u}_a(\bm{x})$ of a through-flow and of an arc-driven flow, both of fixed spatial shape. Measuring lengths in units of $W$ and time in units of $T$, the trajectories then obey
\begin{equation}\label{eq:traj}
\frac{\mathrm{d}\bm{x}}{\mathrm{d}\tilde t}=\frac{1}{\St}\,\bm{u}_b(\bm{x})+\frac{\Lambda}{\St}\,\sin(2\pi\tilde t)\,\bm{u}_a(\bm{x}),
\end{equation}
so that the Lagrangian dynamics depends on two lengths per period: the distance $W/\St$ by which the fluid advances, and the stroke of the forcing, proportional to $\Lambda/\St$. In a creeping flow driven by moving walls alone, the pattern produced by the stirring depends only on the displacement of the walls per period, not on their speed \cite{swanson1990}; the through-flow adds the second length. The superposition can be tested directly on the velocity fields. If it holds, the velocity at every point is a constant plus a sinusoid in phase with the forcing, with the same spatial pattern whatever $\Rey$. It holds closely in the creeping regime. At $\Rey=0.1$, 0.5 and 2, the first harmonic carries 99--100\,\% of the variance of the fluctuating velocity in the bulk; at the maximum of $v_1$ its phase differs from that of the forcing by less than $3^\circ$; its amplitude pattern correlates with that of the reference run with $r\ge0.99$; and $\alpha=0.289$, 0.290 and 0.283. Inertia breaks the superposition. At $\Rey=8$ and 20, the first harmonic carries only 95 and 87\,\% of the variance, its spatial pattern departs from the reference one ($r=0.94$ and 0.74, with local phase shifts of $21^\circ$ and $43^\circ$ rms), and the cross-flow at mid-channel weakens, to $\alpha=0.231$ and 0.162. The stroke, and with it the upper limit of the window derived below, Eq.~\eqref{eq:Stmax}, is then 18 and 43\,\% smaller than in the creeping regime.

\subsection{Two limits and the window between them}\label{subsec:window}
I call stroke the transverse displacement of the fluid at mid-channel during one half period:
\begin{equation}\label{eq:stroke}
S=\frac{\alpha\Omega_0R}{W}\int_0^{T/2}\sin\frac{2\pi t}{T}\dd t=\frac{\alpha\Lambda}{\pi\St}.
\end{equation}
For the two streams to be folded across the channel, the stroke must be at least of the order of the channel width, $S\ge S_c$ with $S_c$ of order one. This is the folding condition of \cite{PhysRevFluids.6.024502}, with the cross-flow of the bulk in place of the wall speed, and it gives an upper bound on the Strouhal number:
\begin{equation}\label{eq:Stmax}
\St\le\St_{\max}=\frac{\alpha\Lambda}{\pi S_c}\simeq0.09\,\Lambda/S_c,
\end{equation}
which is lower than Eq.~\eqref{eq:EOfold} by the factor $\alpha\simeq0.28$: at $\Lambda=27$ and for $S_c=1$, it is 2.4 instead of 8.6. The stroke is $S=2.4$ at the reference forcing, and $S=1.2$ both at $\St=2$ and at $\Lambda=13.5$ with $\St=1$. The map of El Omari et al.\ (their Fig.~13) supports this refinement: on the side of large $\St$ and for $K_0\ge20$, the contours of the outlet standard deviation $\sigma=0.15$ and 0.2 follow approximately lines through the origin, $K_0\approx18\,\St$ and $K_0\approx12$--$14\,\St$, i.e.\ the folding condition with a reduced cross-flow velocity, corresponding to $\alpha\approx0.17$--0.25 for their shallower arcs.

\begin{figure}[!htbp]
\centering
\includegraphics[width=\textwidth]{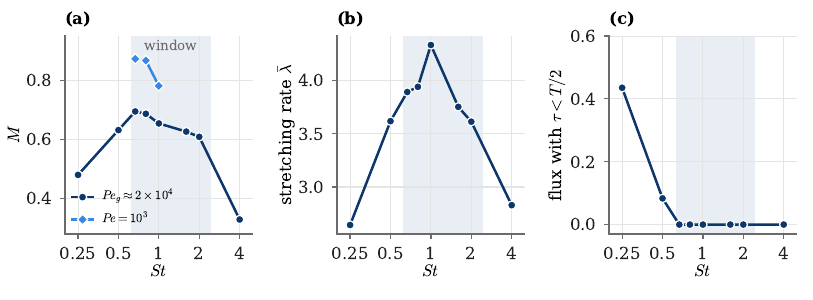}
\caption{Strouhal sweep at $\Lambda=27$. (a) Outlet mixing efficiency, at the grid-limited P\'eclet number (circles) and at $\Pe=10^3$ (diamonds); (b) mean stretching rate of the fluid crossing the mixer; (c) fraction of the flux that crosses the mixing zone in less than half a period, without a reversal of the arcs. Shaded: the window of Eq.~\eqref{eq:window}, $0.62\lesssim\St\lesssim2.4$ ($S_c=1$).}
\label{fig:stsweep}
\end{figure}

The lower limit comes from the other end of the residence-time distribution. A parcel with residence time $\tau$ sees $2\tau/T=2\St\,\tau U_0/W$ reversals of the arcs. A parcel that crosses the mixing zone in less than half a period is carried by a flow whose direction does not change. The separatrices of this flow are barriers to transport (Fig.~2 of \cite{PhysRevFluids.6.024502}), so the parcel is displaced but not folded: it never meets the chaotic saddle, which exists only because the flow reverses. The fraction of the flux in this situation, computed from the residence-time distribution of each run, is zero for $\St\ge0.67$, 0.08 at $\St=0.5$ and 0.44 at $\St=0.25$ (Fig.~\ref{fig:stsweep}c). Every parcel is folded at least once if
\begin{equation}\label{eq:Stmin}
\St\ge\St_{\min}=\frac{W}{2U_0\tau_{\min}}\simeq0.6,
\end{equation}
where $\tau_{\min}\simeq0.40$ varies little for $\St\le1$ (0.38--0.43, Table~\ref{tab:lagr}). This limit expresses, through the fastest fluid rather than the mean residence time, the requirement of the mixing condition of El Omari et al.\ that the fluid see enough reversals of the arcs while it crosses the mixer; it gives a finite Strouhal number, which does not depend on $\Pe$.

Together, the two conditions define a window of efficient periods,
\begin{equation}\label{eq:window}
\frac{W}{2U_0\tau_{\min}}\;\lesssim\;\St\;\lesssim\;\frac{\alpha\Lambda}{\pi S_c},
\end{equation}
whose width, $\St_{\max}/\St_{\min}=\Lambda/\Lambda_c$, grows with the forcing ratio, and which closes at
\begin{equation}\label{eq:Lc}
\Lambda_c=\frac{\pi S_cW}{2\alpha U_0\tau_{\min}}\simeq7\,S_c .
\end{equation}
At $\Lambda=27$ and for $S_c=1$, the window is $0.62\lesssim\St\lesssim2.4$. It leads to three predictions that the simulations can test. First, no period should mix efficiently below $\Lambda_c$. Second, since both limits rise with $\Lambda$, the efficient periods, and presumably the optimum, should move to faster forcing as the forcing ratio increases. Third, since neither limit involves the P\'eclet number, which enters only through the logarithmic number of folds needed to reach the diffusive scale, the range of efficient periods should depend only weakly on $\Pe$, whereas the level of $M$ does depend on it. The window does not say where the optimum lies inside it, nor how high the efficiency is. Both depend on how strongly the saddle stretches the fluid that it does capture, which is the subject of Sec.~\ref{sec:lcs}.

\section{Strouhal number and forcing ratio}\label{sec:StLambda}

\subsection{The optimum period at the reference forcing ratio}\label{subsec:St}
At $\Lambda=27$, the mixing efficiency has a sharp maximum near $\St^*\approx0.7$ (Fig.~\ref{fig:stsweep}a): $M=0.480$, 0.632, 0.695, 0.687, 0.654, 0.626, 0.609 and 0.329 at $\St=0.25$, 0.5, 0.67, 0.8, 1, 1.6, 2 and 4. The maximum is not an effect of the grid-limited P\'eclet number: at $\Pe=10^3$, where the scalar is resolved (Sec.~\ref{subsec:numerics}), it is even more marked, with $M=0.873$ and 0.867 at $\St=0.67$ and 0.8, against 0.781 at $\St=1$ (Fig.~\ref{fig:stsweep}a). The window explains why the efficiency drops on both sides. Between $\St=0.5$ and 0.25, $M$ falls from 0.632 to 0.480 while the fraction of fluid that crosses without a reversal grows from 0.08 to 0.44. Between $\St=2$ and 4, $M$ collapses from 0.609 to 0.329 while the stroke falls from 1.2 to 0.6 channel widths. Inside the window, $M$ varies by less than 0.08. The penetration length of the forcing changes by less than 1\,\% over $\St=0.25$--2, and by 3.5\,\% at $\St=4$ \cite{burghelea_jnnfm}: the period changes the history of the fluid crossing the mixer, not the reach of the stirring.

\begin{figure}[!htbp]
\centering
\includegraphics[width=\textwidth]{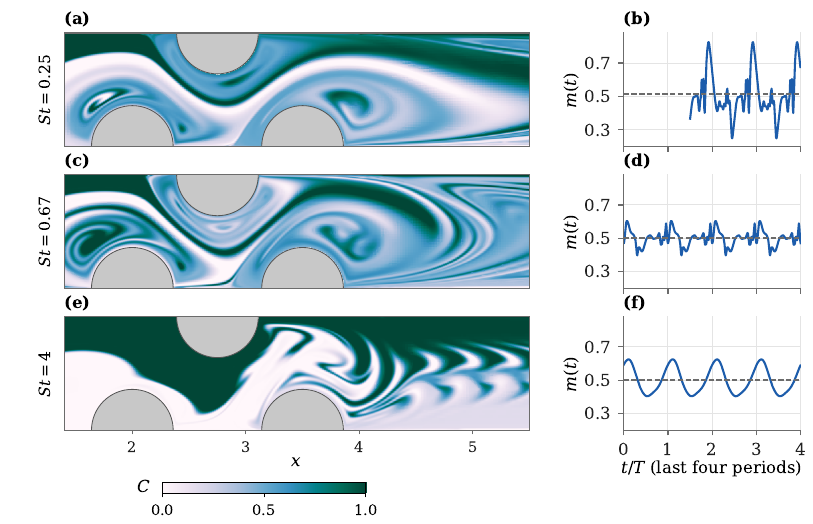}
\caption{Three regimes at $\Lambda=27$, $\Rey=2$: (a, b) $\St=0.25$, below the window; (c, d) $\St=0.67$, at the optimum; (e, f) $\St=4$, above the window. (a, c, e) Concentration field at the start of a forcing period in the periodic regime; grey: arcs. (b, d, f) Composition delivered at the outlet, $m(t)=\int uC\dd y/\int u\dd y$ at $x=5.5$, over the last four periods (two and a half for $\St=0.25$); dashed: time mean.}
\label{fig:regimes}
\end{figure}

Slow and fast forcing fail in different ways, as the concentration fields and the composition delivered at the outlet show (Fig.~\ref{fig:regimes}). At the optimum, the striations of the two streams are folded across the whole channel, and the delivered composition fluctuates by about $\pm0.1$ around its mean. At $\St=0.25$, the streams are drawn into long filaments, but part of the fluid is swept through between two reversals and leaves in a pulse once per period. At $\St=4$, the arcs reverse before the fluid near mid-channel has moved by one channel width; the streams are folded only near the arcs and leave the mixer largely segregated, with a weaker pulsation. Splitting the outlet variance into its spatial and temporal parts quantifies this contrast. The temporal part is a quarter of the variance at $\St=0.25$ and 0.5, against 9\,\% at the optimum, and it accounts for two thirds of the increase in variance between $\St=0.67$ and 0.5. With fast forcing, the temporal part is only 3--9\,\% of the variance, but the spatial part at $\St=4$ is five times larger than at the optimum. Slow forcing thus makes the outlet composition vary in time, and fast forcing makes it vary across the section.

The optimum is not where the fluid is stretched most. The mean stretching rate of the fluid crossing the mixer (Fig.~\ref{fig:stsweep}b) is largest at $\St=1$ (4.34), not at the optimum (3.90 at $\St=0.67$), and it decreases on both sides, to 2.64 at $\St=0.25$ and 2.83 at $\St=4$. Between $\St=0.67$ and 1 the stretching rate still increases by 11\,\%, yet the outlet variance increases by 0.0070, and 0.0048 of this increase comes from the pulsation (0.0081 and 0.0042 at $\Pe=10^3$). At $\St=1$ the delivered composition locks onto the forcing: 74\,\% of its variance is in the first harmonic, against 22\,\% at $\St=0.67$. At the optimum, the fluid is therefore not the most stretched, but it is delivered most steadily; the optimum is the slowest sampled period at which every parcel is reversed at least once (the unreversed fraction is 0.08 at $\St=0.5$ and zero from $\St=0.67$). The experiments \cite{kiwan2025} also found the maxima of the stretching and of the mixing at different Strouhal numbers, with the maximum of the mixing at the same Strouhal number as here ($\St\approx0.7$). Their stretching maximum, however, lies at slower forcing ($\St\approx0.4$), and it was obtained with a finite-time Lyapunov exponent integrated over a fixed time of 0.25\,s; with an integration time proportional to the period, the measured exponent increased monotonically with $\St$. The location of a stretching maximum thus depends on the time horizon over which the stretching is measured, whereas the stretching of the fluid crossing the mixer used here has none.

\subsection{The forcing ratio and the shift of the optimum}\label{subsec:Lambda}
\begin{figure}[!htbp]
\centering
\includegraphics[width=0.62\textwidth]{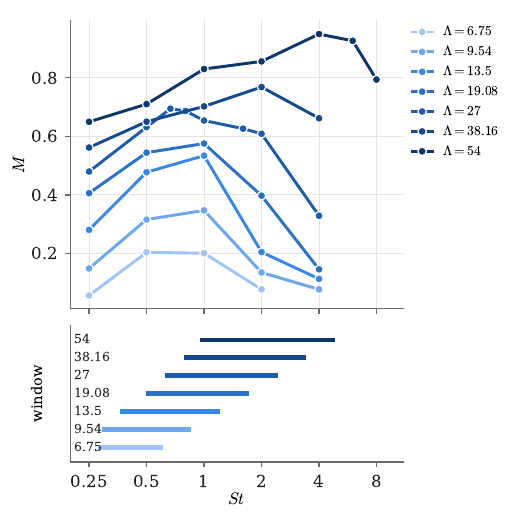}
\caption{Outlet mixing efficiency versus the Strouhal number for the seven forcing ratios (top), and the window of Eq.~\eqref{eq:window} for each forcing ratio, with $\tau_{\min}$ measured at $\St=1$ (bottom).}
\label{fig:map}
\end{figure}

Stronger forcing mixes better at every period. At $\St=1$, $M=0.201$, 0.347, 0.534, 0.576, 0.654, 0.702 and 0.830 for $\Lambda=6.75$, 9.54, 13.5, 19.08, 27, 38.16 and 54 (Fig.~\ref{fig:map}). As predicted, no period mixes well below the critical forcing ratio: at $\Lambda=6.75$, $M=0.057$, 0.204, 0.201 and 0.077 at $\St=0.25$, 0.5, 1 and 2, against 0.53 at $\Lambda=13.5$. The estimate $\Lambda_c\approx7$ uses the value of $\tau_{\min}$ measured at $\Lambda=27$. With weaker forcing the fastest fluid is slower, because the arcs carry it more slowly ($\tau_{\min}=0.90$ at $\Lambda=6.75$). A narrow window, $0.28\lesssim\St\lesssim0.61$, therefore remains at $\Lambda=6.75$, and it contains the best of the four periods tested. Just above $\Lambda_c$, at $\Lambda=9.54$, the window is $0.29\lesssim\St\lesssim0.86$, yet the efficiency reaches only 0.35, because the stretching inside the window is weak (1.70 at $\St=0.5$, against 3.62 at $\Lambda=27$). Being inside the window is necessary for efficient mixing, but not sufficient.

As the forcing becomes stronger, the window widens and moves to faster periods. Both limits increase with $\Lambda$. The upper limit grows as $\alpha\Lambda/\pi$. The lower limit grows because the fastest fluid, carried by the arcs, crosses the mixer sooner when the arcs turn faster: $\tau_{\min}=0.86$, 0.69, 0.50, 0.40, 0.32 and 0.26 at $\Lambda=9.54$, 13.5, 19.08, 27, 38.16 and 54, so that the window is 0.29--0.86, 0.36--1.2, 0.50--1.7, 0.62--2.4, 0.78--3.4 and 0.95--4.9. In each series, the efficiency is high inside the window and drops outside it (Fig.~\ref{fig:map}). Inside the window, $M$ is within 0.12 of the best sampled value at the same forcing ratio. Outside it, $M$ is lower by 0.06--0.43, except at three points just beyond the upper limit, where it is within 0.02 of the best value: $\St=1$ at $\Lambda=6.75$ and 9.54 ($\St_{\max}=0.61$ and 0.86), and $\St=6$ at $\Lambda=54$ ($\St_{\max}=4.9$). The upper limit is thus less sharp than the lower one, as expected for a condition on a stroke of order one channel width. The window therefore locates the efficient periods over a factor of eight in $\Lambda$.

The window contains one constant, $S_c$, and one threshold, the quantile $F=10^{-3}$ that defines $\tau_{\min}$. The threshold hardly matters: for $F=10^{-4}$ to $10^{-2}$, $\St_{\min}=0.59$--0.62 at $\Lambda=27$ and $\Lambda_c=6.5$--6.9. The constant is constrained by the data. Calling a run efficient when its $M$ is within 0.05 of the best sampled value at its forcing ratio, the window agrees with this classification for 30 of the 39 runs when $0.7\le S_c\le1.2$, against 28 for $S_c=0.5$ and 26 for $S_c=2$; with $S_c=2$, it contains the best sampled period at only one of the seven forcing ratios. The value $S_c=1$ used throughout lies within the range allowed by the data.

\begin{figure}[!htbp]
\centering
\includegraphics[width=\textwidth]{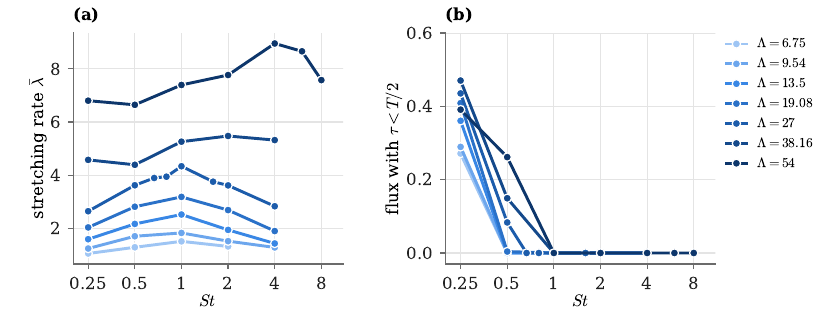}
\caption{Lagrangian measures versus the Strouhal number for the seven forcing ratios: (a) mean stretching rate of the fluid crossing the mixer; (b) fraction of the flux that crosses the mixing zone in less than half a period.}
\label{fig:lambdalagr}
\end{figure}

The optimum follows the window to faster forcing, but its position inside the window changes. The best sampled period is $\St=0.5$ at $\Lambda=6.75$, 1 at $\Lambda=9.54$--19.08, 0.67 at $\Lambda=27$, 2 at $\Lambda=38.16$ and 4 at $\Lambda=54$. At $\Lambda=54$ the optimum is bracketed between $\St\approx4$ and 6, at the upper limit of the window ($\St_{\max}=4.9$). At $\Lambda=27$, the only forcing ratio sampled more finely than by factors of two, the optimum lies near the lower limit. At $\Lambda\le19.08$ the sampling is too coarse to locate it: on the same factor-of-two grid, the series at $\Lambda=27$ would also give $\St=1$. The optimum moves up because, with strong forcing, even a single stroke stretches the fluid strongly (Fig.~\ref{fig:lambdalagr}). The stretching rate grows with the forcing ratio, from 1.4--2.5 at $\Lambda=13.5$ to 4.4--5.5 at $\Lambda=38.16$ and 6.6--9.0 at $\Lambda=54$. At $\Lambda=54$ it increases with $\St$ up to $\St=4$ (8.96), as $M$ does, and then decreases (8.66 and 7.58 at $\St=6$ and 8). The unreversed fraction of the flux vanishes at the lower limit of the window for every $\Lambda$, as assumed in Eq.~\eqref{eq:Stmin}. With strong forcing, however, the fluid that escapes without a reversal is already well stretched: at $\St=0.5$, 15\,\% of the flux is unreversed at $\Lambda=38.16$ and 26\,\% at $\Lambda=54$, and yet $M=0.650$ and 0.710. The reversal condition thus matters most when the stretching is weak. When the stretching is strong, the mixing improves with the number of strokes per transit until the strokes become too short, and the optimum moves towards the upper limit of the window.

\begin{figure}[!htbp]
\centering
\includegraphics[width=0.62\textwidth]{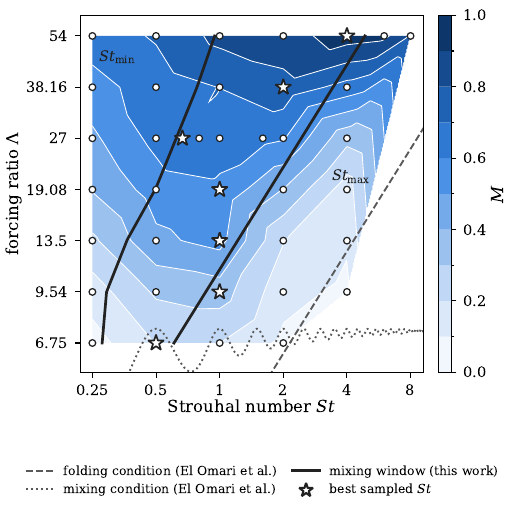}
\caption{Efficiency map: outlet mixing efficiency in the $(\St,\Lambda)$ plane ($\Rey=2$, $\Pe=2\times10^7$, level 10), interpolated linearly in $(\log\St,\log\Lambda)$ between the simulations (dots); stars: largest sampled efficiency at each forcing ratio. Solid lines: limits of the window, Eq.~\eqref{eq:window}, with $\tau_{\min}$ measured at $\St=1$ for each forcing ratio. Dashed and dotted lines: folding and mixing conditions of El Omari et al., Eqs.~\eqref{eq:EOfold} and~\eqref{eq:EOmix} ($\zeta=2$, $\Pe=10^6$, as in \cite{PhysRevFluids.6.024502}).}
\label{fig:emap}
\end{figure}

The 39 simulations at $\Rey=2$ are combined in Fig.~\ref{fig:emap} into a map of the efficiency in the $(\St,\Lambda)$ plane, in the form used by El Omari et al.\ \cite{PhysRevFluids.6.024502} (their Fig.~13, 90 simulations at $\Pe=10^6$). The region of efficient mixing ($M>0.6$) widens towards large forcing ratios between the two limits of the window, which meet near $\Lambda\approx7$. The mixing condition of El Omari et al., Eq.~\eqref{eq:EOmix}, gives a similar minimum forcing ratio. Their folding condition, Eq.~\eqref{eq:EOfold}, has the form of the upper limit of the window, a line through the origin; with the wall speed as the velocity of the cross-flow it lies at faster forcing, and the measured cross-flow at mid-channel brings it onto the edge of the efficient region. The map is sampled at seven forcing ratios, in steps of about $\sqrt2$, and at four to eight periods, mostly in steps of two. The optimum is bracketed at strong forcing, but at weaker forcing its position inside the window is known only to within a factor of two.

\section{Reynolds number}\label{sec:Re}
\begin{figure}[!htbp]
\centering
\includegraphics[width=0.5\textwidth]{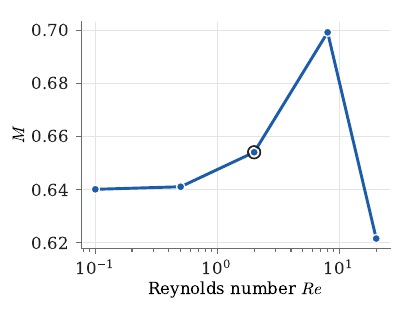}
\caption{Outlet mixing efficiency at $\St=1$, $\Lambda=27$, $\Pe=2\times10^7$ versus the Reynolds number. The reference run ($\Rey=2$) is circled. The dependence on the P\'eclet number is discussed in Sec.~\ref{sec:Pe}.}
\label{fig:re}
\end{figure}

In the creeping regime the efficiency hardly depends on the Reynolds number: at $\St=1$ and $\Lambda=27$, $M=0.640$, 0.641 and 0.654 at $\Rey=0.1$, 0.5 and 2 (Fig.~\ref{fig:re}). The reference runs are therefore representative of the creeping flow of the experiments. Moderate inertia appears to improve the mixing ($M=0.699$ at $\Rey=8$, a gain checked at one resolution only, Sec.~\ref{subsec:numerics}), and stronger inertia degrades it ($M=0.622$ at $\Rey=20$). The flow remains time-periodic: at $\Rey=8$ and 20, as at the reference, the efficiency computed period by period varies by less than 0.004. The stretching alone does not explain this trend, since the mean stretching rate of the fluid crossing the mixer keeps increasing with inertia: $\bar\lambda=3.94$ and 3.84 at $\Rey=0.1$ and 0.5, and 4.34, 4.57 and 5.06 at $\Rey=2$, 8 and 20. Neither does the confinement of the forcing observed with shear-thinning fluids \cite{burghelea_jnnfm}: the penetration length also decreases monotonically, from 0.0648 at $\Rey=0.1$ to 0.0637, 0.0573 and 0.0530 at $\Rey=2$, 8 and 20. Instead, the highest $M$, at $\Rey=8$, coincides with the smallest fraction of poorly stretched fluid ($s<2$: 0.24, 0.22, 0.15, 0.10 and 0.15 across the series) and with the largest median log-stretch (4.5, 4.6, 5.6, 5.7 and 5.7). At $\Rey=20$ the stretching is the strongest of the series, but the mean residence time is 9\,\% shorter (1.39 against 1.51--1.54) and the poorly stretched fraction rises again. Inertia thus stirs harder but reaches less of the fluid. One reason is that the cross-flow at mid-channel weakens ($\alpha=0.231$ at $\Rey=8$ and 0.162 at $\Rey=20$, against 0.283; Sec.~\ref{subsec:linear}), so that at $\Rey=20$ the reference period lies close to the upper limit of the window ($S=1.4$). The other reason, a weakly stirred wake, appears in the Lagrangian structure (Sec.~\ref{subsec:lcsRe}). Note finally that the wall Reynolds number, $\Rey_w=\Lambda\Rey$, ranges from 2.7 to 540 across the series: even at the reference value $\Rey=2$, the flow near the arcs is not a creeping flow, although the through-flow is.

\section{How the control parameters reshape the chaotic saddle and its coherent structures}\label{sec:lcs}

Three results of Secs.~\ref{sec:StLambda} and \ref{sec:Re} are not explained by the kinematic window: with fast forcing, the efficiency collapses although the fluid is still stretched at a substantial rate; with strong forcing, the optimum moves to the upper limit of the window; and inertia degrades the mixing while it strengthens the stretching. The structure of the chaotic saddle, computed for the three one-parameter series through the reference point (Figs.~\ref{fig:manst}--\ref{fig:rtd}, Table~\ref{tab:lagr}), explains all three. In every run, elliptic islands cover at most 1.4\,\% of the area (0.7\,\% for $\Lambda\ge19$), so that the flow is globally chaotic, and the manifolds cross the channel in three to six folds per section. The parameters change the way the stretching is shared between the ridges and the regions between them more than the topology of the manifolds.

\subsection{Strouhal number: lobes of a nearly steady flow, and retention without folding}\label{subsec:lcsSt}
Inside the window, the saddle hardly changes. From $\St=0.5$ to 2, its Lyapunov exponent is 4.1--5.4, its escape rate 1.0--1.25 and the dimension of its unstable manifold 1.73--1.81 (Fig.~\ref{fig:chaosstats}d,g). The manifolds form a dense web of thin filaments across the mixing zone (Fig.~\ref{fig:manst}c--h), and the contrast of the stretching field is at its lowest, $H=1.40$--1.52, with a minimum between the optimum and the reference period. At a given $\Lambda$ the saddle is thus a robust feature of the forcing, and the differences in efficiency inside the window come from the way the fluid is delivered to it, as shown in Sec.~\ref{subsec:St}.

Outside the window, the structure is deformed in two opposite ways. With slow forcing ($\St=0.25$), the stroke, about ten channel widths per half period, is long compared with the distance the fluid advances, and the flow is almost steady during each half period. Its manifolds are long lobes that follow the instantaneous streamlines (Fig.~\ref{fig:manst}a,b); the FTLE fields show large, weakly stretched cells around the arcs, and a lobe of slowly rotating fluid downstream of the last arc, where the long-lived tracers gather. The saddle is weak ($\lambda_s=2.83$), although it holds the fluid longer than at the reference ($\kappa=0.80$). The residence-time distribution is the broadest of the series ($c_v=0.73$), with an early peak at $\tau\approx0.5$, made of fluid swept through between two reversals, and a long tail of fluid that turns in the lobes. The contrast is high ($H=1.70$), and 44\,\% of the flux crosses with $s<2$. Slow forcing thus fails because the saddle is fed only once per half period; in between, the fluid follows the cells of a steady flow.

With fast forcing ($\St=4$), the stroke is shorter than the channel width. The arcs reverse before they can carry fluid along their surface, and the fluid crosses the mixer as a nearly uniform stream: the residence-time distribution is narrow, with no fast fluid ($\tau_{\min}=0.94$). Yet the saddle holds the fluid longer than at any other period: $\kappa=0.35$, and 8\,\% of the tracers remain after four time units, against 2\,\% at the reference (Fig.~\ref{fig:chaosstats}a), most of them in the bulk near the upstream arcs, where the short strokes shake the fluid back and forth faster than they fold it. The forward FTLE shows dense, short ridges near the arcs, but also a weakly stretched region downstream of the top arc, which covers 30\,\% of the area behind the last arc, against 15\,\% at the reference (Fig.~\ref{fig:manst}i,j). The saddle stretches at only $\lambda_s=3.05$, and yet its unstable manifold has the largest dimension of the series, $D_u=1.88$, because the escape is slow. The dimension of the unstable manifold, often used to measure the efficiency of chaotic transport in open flows \cite{jung1993,lt2011}, is therefore misleading here: within the Strouhal series, the efficiency increases with the escape rate ($r=0.93$) and decreases with $D_u$ ($r=-0.65$). Holding the fluid helps only if the fluid is stretched while it is held.

\subsection{Forcing ratio: a saddle that grows to fill the mixing zone}\label{subsec:lcsLambda}
\begin{figure}[!htbp]
\centering
\includegraphics[width=\textwidth]{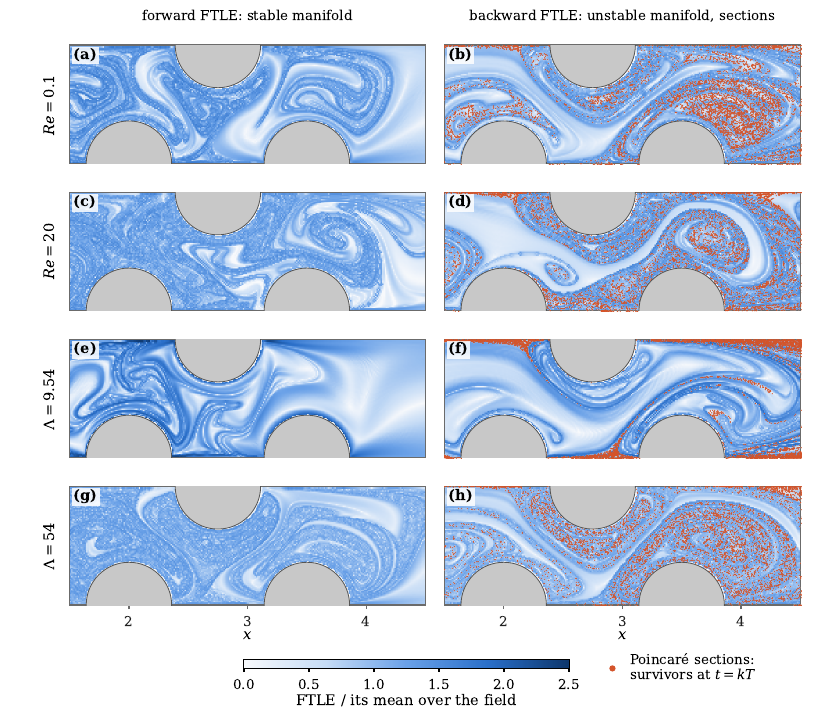}
\caption{As Fig.~\ref{fig:manst}, at $\St=1$ for (a, b) $\Rey=0.1$ and (c, d) $\Rey=20$ ($\Lambda=27$), and for (e, f) $\Lambda=9.54$ and (g, h) $\Lambda=54$ ($\Rey=2$).}
\label{fig:manrl}
\end{figure}

The forcing ratio acts on the strength of the saddle rather than on the delivery of fluid to it, and its effect is monotonic. From $\Lambda=6.75$ to 54 at $\St=1$, the Lyapunov exponent of the saddle rises from 1.91 to 6.53, the dimension of the unstable manifold from 1.48 to 1.88, and the contrast falls from 1.97 to 1.25 (Fig.~\ref{fig:chaosstats}f,i,l); the weakly stretched area shrinks from 19\,\% to 6\,\% of the mixing zone, and the small islands disappear. With weak forcing, the saddle is a thin set of filaments near the arcs (Fig.~\ref{fig:manrl}e,f), and the fluid crosses the mixer through a wide, weakly stretched region downstream of the top arc. The long-lived fluid is then not on the saddle but along the walls (40--48\,\% of the survivors after two time units at $\Lambda\le13.5$, against 19\,\% at the reference), and the residence-time distribution is narrow ($c_v\approx0.3$), close to that of a perturbed plug flow. With strong forcing (Fig.~\ref{fig:manrl}g,h), the manifolds and the Poincar\'e sections fill the bulk, and the arcs draw fluid into large lobes that recirculate between them. The residence-time distribution broadens ($c_v=0.69$ at $\Lambda=54$), but the long-lived fluid is held in the bulk, where it is stretched (11\,\% of the survivors are near a wall). The escape rate rises from 1.0 to 1.7 between $\Lambda=6.75$ and 13.5, as the arcs sweep fluid out of the near-wall regions, and falls to 0.8 at $\Lambda\ge38$, as the lobes hold it. This is why the optimum moves to faster forcing at strong forcing ratios: once the saddle fills the zone, each additional stroke adds stretching to fluid that is already near it, and short residence times no longer matter.

The scaling of Eq.~\eqref{eq:traj} shows up in the residence times. For strokes shorter than about two channel widths, $\tau_{\min}$ depends only on the stroke, whether it is shortened by fast or by weak forcing: 0.94 and 0.90 for $S=0.61$ ($\St=4$, $\Lambda=27$ and $\St=1$, $\Lambda=6.75$), 0.72 and 0.69 for $S=1.22$ ($\St=2$, $\Lambda=27$ and $\St=1$, $\Lambda=13.5$). For longer strokes, $\tau_{\min}$ levels off at about 0.40 as the period increases, but keeps decreasing as $\Lambda$ increases, because the fluid carried by the arcs moves at a speed proportional to $\Lambda$. The saddle does not follow the stroke in the same way ($\lambda_s=3.05$ and 1.91 at $S=0.61$): for the same stroke, a faster wall stretches more.

\subsection{Reynolds number: a stronger saddle and a weakly stirred wake}\label{subsec:lcsRe}
Inertia has two effects on the structure of the flow. In the core of the mixing zone, it strengthens the saddle and makes the stretching more uniform: $\lambda_s$ increases from 4.07 at $\Rey=0.1$ to 6.16 at $\Rey=20$, and $H$ falls from 1.50 to 1.29 (Fig.~\ref{fig:chaosstats}e,k). In the creeping regime (Fig.~\ref{fig:manrl}a,b), the manifolds are almost those of the reference run, and the saddle holds the fluid longer ($\kappa=0.71$--0.77, against 1.25 at $\Rey=2$); this longer retention presumably compensates for the weaker stretching, so that $M$ hardly changes between $\Rey=0.1$ and 2. At $\Rey=20$ (Fig.~\ref{fig:manrl}c,d), a large vortex forms downstream of the top arc, and a weakly stirred wake extends behind it: the weakly stretched area downstream of the last arc grows from 15\,\% at $\Rey=2$ to 28\,\% at $\Rey=8$ and 33\,\% at $\Rey=20$, the survivors shift upstream, and the mean residence time falls by 9\,\%. At $\Rey=8$ the stronger saddle wins, and the poorly stretched fraction is the smallest of all the runs at $\Lambda=27$ (0.10); at $\Rey=20$ the wake and the shorter transit win, and it rises again to 0.15. Inertia thus strengthens the saddle, but also creates a coherent region inside the mixing zone that the saddle does not reach.

\subsection{What the Lagrangian structure tells about the mixing}\label{subsec:lcsM}
\begin{figure}[!htbp]
\centering
\includegraphics[width=\textwidth]{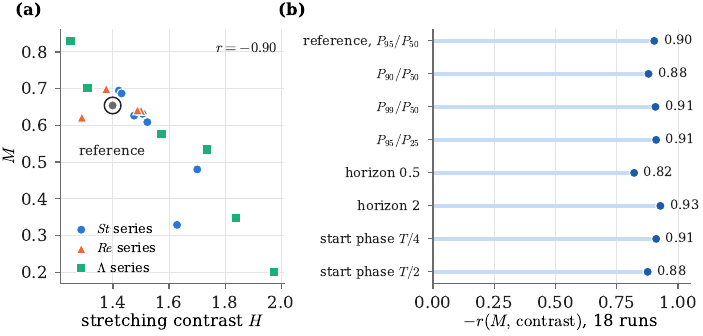}
\caption{Mixing efficiency and stretching contrast for the 18 runs of Table~\ref{tab:lagr}. (a) $M$ versus $H$, the ratio of the 95th percentile to the median of the forward FTLE over one time unit from phase 0; circled: reference run. (b) Correlation coefficient between $M$ and the contrast over the 18 runs, for the reference definition and for other percentiles $P_q$ of the forward FTLE, other horizons (in time units) and other start phases.}
\label{fig:mh}
\end{figure}

The three series show that no single property of the saddle sets the mixing efficiency. Across the 18 runs of Table~\ref{tab:lagr}, $M$ correlates with the Lyapunov exponent of the saddle ($r=0.85$), but this correlation comes mostly from the forcing-ratio series ($r=0.98$) and vanishes in the Reynolds series ($r=0.01$). $M$ correlates only weakly with the dimension of the unstable manifold ($r=0.64$), and not at all with the escape rate. The best correlates all measure the weakly stretched fluid: the contrast of the stretching field ($r=-0.90$), the median log-stretch of the fluid crossing the mixer ($r=0.89$) and its poorly stretched fraction ($r=-0.83$). Of these, the contrast $H$, Eq.~\eqref{eq:H}, is the most useful, because it is a property of the velocity field alone: it is computed over one time unit, without following the fluid through the whole mixer. $H$ also correlates with $M$ within each series ($r=-0.84$ for $\St$ and $-0.97$ for $\Lambda$), except in the Reynolds series, where the efficiency varies by only 0.08 and the loss at $\Rey=20$ occurs in the wake, downstream of most of the stretching. Figure~\ref{fig:mh}a shows the relation for all 18 runs. It does not depend on the details of the definition of $H$ (Fig.~\ref{fig:mh}b): with the 90th or 99th percentile instead of the 95th, the 25th percentile instead of the median, a horizon of 0.5 or 2 time units, or a start at a quarter or a half of the period, $r=-0.82$ to $-0.93$. The shortest horizon degrades the correlation most, mainly in the Strouhal series ($r=-0.46$), because at slow forcing half a time unit is less than half a period; with a horizon of two time units, $H$ also follows $M$ in the Reynolds series ($r=-0.77$). These correlations are computed over 18 runs from three series that share the reference run, not on independent data: they identify the property of the flow that follows the efficiency, not a predictive law. In this open flow the mixing is limited by the least stirred region: the cells of an almost steady flow when the forcing is slow, the regions that short strokes cannot fold when it is fast, the near-wall regions and the unstirred passage when it is weak, and the wake when inertia is large. Each control parameter acts on the mixing through the coherent structure that it creates or removes.

\section{P\'eclet number}\label{sec:Pe}
\begin{figure}[!htbp]
\centering
\includegraphics[width=0.6\textwidth]{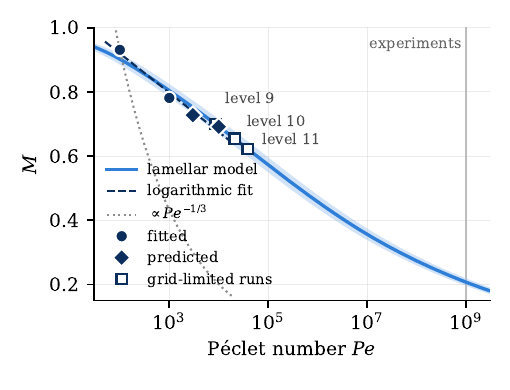}
\caption{Outlet mixing efficiency versus P\'eclet number at $\St=1$, $\Lambda=27$. Circles: runs used to fit the initial striation thickness $\delta_0$ of the lamellar model (solid line; band: $\delta_0$ varied by $\pm25\,\%$); diamonds: runs at $\Pe=3\times10^3$ and $10^4$, predicted by the model; dashed: logarithmic fit; dotted: $\Pe^{-1/3}$ through the run at $\Pe=10^2$. Squares: runs at the nominal $\Pe=2\times10^7$ at levels 9--11, placed at the effective P\'eclet number of the grid obtained by inverting the model. The vertical line marks the P\'eclet number of the experiments \cite{kiwan2025}.}
\label{fig:pescaling}
\end{figure}

All the efficiencies discussed so far were computed at the effective P\'eclet number of the grid. In Eckart's description of stirring \cite{eckart1948,branicki2010}, the deformation of the fluid first builds up sharp gradients, which diffusion then removes; the P\'eclet number sets the scale at which the second stage takes over from the first. At $\St=1$ and $\Lambda=27$, the efficiency decreases from $M=0.931$ at $\Pe=10^2$ to 0.781 at $10^3$, 0.728 at $3\times10^3$ and 0.692 at $10^4$ (Fig.~\ref{fig:pescaling}). This is much slower than the power laws of laminar mixing without chaos. If the striations were thinned at a constant rate, as in a shear flow, the diffusive length at the outlet would scale as $\Pe^{-1/3}$ (L\'ev\^eque scaling), and $M$ would be halved for each decade of $\Pe$. Over these two decades, a power law with a free exponent, $M\propto\Pe^{-0.066}$, and a logarithmic law, $M=1.165-0.123\log_{10}\Pe$, fit the data about equally well (rms deviations 0.010 and 0.014), and neither is exact: the local exponent of the power law decreases in magnitude from $-0.076$ to $-0.042$ over the three intervals. Such a weak dependence is the signature of exponential stretching. A striation of initial thickness $\delta_0$ stretched by a factor $e^s$ reaches the diffusive scale when $s\simeq\tfrac12\ln\Pe$, so a tenfold increase in $\Pe$ requires only about one more e-fold of stretching.

The measured stretching makes this argument quantitative. Each parcel crossing the mixer is stretched by a factor $e^{s}$ during its residence time $\tau$. I treat it as a lamella of initial thickness $\delta_0$ stretched at the constant rate $\gamma=s/\tau$, which, by incompressibility, compresses it at the same rate in the normal direction. In a frame attached to the lamella, with $x'$ the coordinate normal to it, the velocity across the lamella is $-\gamma x'$, and the concentration obeys \cite{ranz1979,villermaux2019}
\begin{equation}
\frac{\partial C}{\partial t}-\gamma x'\frac{\partial C}{\partial x'}=\mathcal{D}\,\frac{\partial^2C}{\partial x'^2}.
\label{eq:lamella}
\end{equation}
The variables $\xi=x'e^{\gamma t}/\delta_0$, which follows the compression, and $\theta(t)=(\mathcal{D}/\delta_0^2)\int_0^te^{2\gamma t'}\dd t'$, a time weighted by the square of the stretching, remove the advection term and turn Eq.~\eqref{eq:lamella} into the diffusion equation $\partial C/\partial\theta=\partial^2C/\partial\xi^2$. For an initial striation $C-\tfrac12\propto e^{-\xi^2}$, its solution is $C-\tfrac12\propto(1+4\theta)^{-1/2}\exp[-\xi^2/(1+4\theta)]$: the concentration contrast of the lamella is reduced by the factor $(1+4\theta)^{-1/2}$, with, at the outlet ($t=\tau$),
\begin{equation}
\theta=\frac{\mathcal{D}\tau}{\delta_0^2}\,\frac{e^{2s}-1}{2s}.
\label{eq:ranz}
\end{equation}
Without stretching ($s\to0$), $\theta=\mathcal{D}\tau/\delta_0^2$, the residence time measured in units of the diffusion time across a striation of fixed thickness. With strong stretching, $\theta\simeq(\mathcal{D}\tau/\delta_0^2)\,e^{2s}/(2s)$: stretching multiplies the effect of diffusion by $e^{2s}$. A parcel is mixed when $\theta$ reaches order one; with $\mathcal{D}=U_0W/\Pe$, this happens when $s\simeq\ln[(\delta_0/W)\Pe^{1/2}]$, up to a logarithmic correction, as anticipated above. At the inlet, the two streams carry $C=0$ and 1, so that $\sigma_{\mathrm{tot}}=\tfrac12$ and $M=0$. Approximating the outlet standard deviation by the sum of the residual contrasts of the parcels, each weighted by the fraction $w_i$ of the flux it carries, gives $\sigma_{\mathrm{tot}}=\tfrac12\sum_iw_i(1+4\theta_i)^{-1/2}$, i.e.\ $M=1-\sum_iw_i(1+4\theta_i)^{-1/2}$. The parcels are treated as independent lamellae; their merging, which slows the decay of the variance in well-mixed regions \cite{villermaux2019}, is ignored. The model has a single parameter, $\delta_0$. Fitted to the runs at $\Pe=10^2$ and $10^3$, it gives $\delta_0=0.23W$, about a quarter of the channel width. With this value, the model predicts $M=0.752$ at $\Pe=3\times10^3$ and 0.692 at $10^4$, against 0.728 and 0.692 in the simulations (Fig.~\ref{fig:pescaling}, diamonds). It misses the two fitted points by 0.03 and 0.02, because the stretching rate is not constant along a trajectory. The slope $-\mathrm{d}M/\mathrm{d}\log_{10}\Pe$ of the model increases from 0.08 at $\Pe=10^2$ to 0.12 at $10^4$--$10^5$ and then decreases slowly, as the least stretched parcels become the only ones left unmixed. The logarithmic law is therefore only a local approximation: $M$ is essentially the fraction of the flux stretched beyond $s\simeq\ln(\delta_0\Pe^{1/2})$. This reinforces the conclusion of Sec.~\ref{sec:lcs}: the higher the P\'eclet number, the more the efficiency depends on the poorly stretched tail of the distribution.

Inverting the model for the runs at the nominal $\Pe=2\times10^7$ gives the effective P\'eclet number set by numerical diffusion: $8.6\times10^3$, $2.1\times10^4$ and $3.9\times10^4$ at levels 9, 10 and 11 (Fig.~\ref{fig:pescaling}, squares). A joint fit of $\delta_0$ and of a grid diffusivity added to the molecular one, on the four runs at level 10, gives the same value, $\Pe_g=2.0\times10^4$ (with $\delta_0=0.25W$). The numerical diffusivity thus decreases roughly in proportion to the cell size, as expected when the scalar fronts are not resolved. It follows that the run at $\Pe=10^4$ is itself slightly affected by the grid: the model gives $M=0.705$ with the combined diffusivity, against 0.685 with the molecular one alone. Repeated at level 11, this run gives $M=0.678$ instead of 0.692. The change, 0.014, is less than half of that at the nominal P\'eclet number (0.032), since molecular diffusion now dominates the numerical one, and the value at level 11 is already close to the molecular estimate. It also follows that the efficiencies of Secs.~\ref{sec:StLambda} and \ref{sec:Re} are those of a fluid with $\Pe\approx2\times10^4$. Comparisons between runs at the same resolution remain meaningful, since the effective diffusivity is a property of the grid, but at the P\'eclet numbers of the experiments ($\sim10^9$) the model gives $M\approx0.2$ at the reference forcing.

Applied without any change ($\delta_0=0.23W$, $\Pe_g=2.1\times10^4$) to the 39 runs of the $(\St,\Lambda)$ map, the model reproduces the simulated efficiencies with $r^2=0.93$ (rms deviation 0.072, maximum 0.14), and it finds the best sampled period at six of the seven forcing ratios. Its errors are systematic, and they are informative. The model overestimates $M$ with fast forcing outside the window (0.44 against 0.33 at $\St=4$, $\Lambda=27$; 0.29 against 0.15 at $\St=4$, $\Lambda=19.08$), and underestimates it with slow forcing at $\Lambda=27$ (by 0.06--0.10 for $\St\le0.8$). As a result, it places the optimum at $\Lambda=27$ at the maximum of the stretching rate, $\St=1$, instead of $\St\approx0.7$, where the simulations place it both at $\Pe_g$ and at $\Pe=10^3$. A model that knows only how much each parcel is stretched cannot see how the striations of the two streams are arranged across the section and in time, which is precisely what distinguishes slow from fast forcing (Sec.~\ref{subsec:St}). It also underestimates $M$ at $\Rey\le0.5$ (0.54--0.56 against 0.64), where it does not account for the longer retention in the saddle. With these caveats, the model shows how the map changes with $\Pe$ (Table~\ref{tab:pemap}). With strong forcing, the best sampled period changes little between $\Pe_g$ and $10^9$, as the window predicts; with weak forcing, it moves to slower periods as $\Pe$ increases. More importantly, the efficiency falls much faster with weak forcing than with strong forcing. At $\Pe=10^9$, the best efficiency is 0.02, 0.05, 0.10, 0.21, 0.32 and 0.59 at $\Lambda=9.54$, 13.5, 19.08, 27, 38.16 and 54, because with weak forcing a large fraction of the flux is poorly stretched (Table~\ref{tab:lagr}). The forcing ratio below which the mixer is useless therefore increases with the P\'eclet number. This trend is captured by the mixing condition of El Omari et al., whose minimum forcing ratio grows as $\ln\Pe$ (5.2, 7.4 and 11 at $\Pe=2\times10^4$, $10^6$ and $10^9$ for $\zeta=2$, Eq.~\eqref{eq:EOmix}); the window, Eq.~\eqref{eq:window}, which addresses only the kinematics, closes at $\Lambda_c\approx7$ whatever the P\'eclet number. The two conditions are complementary.

\begin{table}[!htbp]
\centering
\caption{Best sampled Strouhal number and corresponding mixing efficiency for each forcing ratio: simulations (grid-limited, $\Pe_g\approx2\times10^4$) and lamellar model at $\Pe_g$, $10^6$ and $10^9$.}
\label{tab:pemap}
\begin{tabular}{@{}ccccc@{}}
\toprule
$\Lambda$ & simulations & model, $\Pe_g$ & model, $10^6$ & model, $10^9$\\
\midrule
6.75 & 0.5 (0.204) & 1 (0.213) & 0.5 (0.066) & 0.25 (0.013)\\
9.54 & 1 (0.347) & 1 (0.296) & 0.5 (0.120) & 0.5 (0.024)\\
13.5 & 1 (0.534) & 1 (0.450) & 1 (0.187) & 0.5 (0.048)\\
19.08 & 1 (0.576) & 1 (0.547) & 1 (0.326) & 1 (0.098)\\
27 & 0.67 (0.695) & 1 (0.654) & 1 (0.457) & 1 (0.207)\\
38.16 & 2 (0.768) & 2 (0.800) & 2 (0.605) & 1 (0.315)\\
54 & 4 (0.949) & 4 (0.926) & 4 (0.830) & 4 (0.591)\\
\bottomrule
\end{tabular}
\end{table}

\section{Energy cost and operating map}\label{sec:energy}
\begin{figure}[!htbp]
\centering
\includegraphics[width=\textwidth]{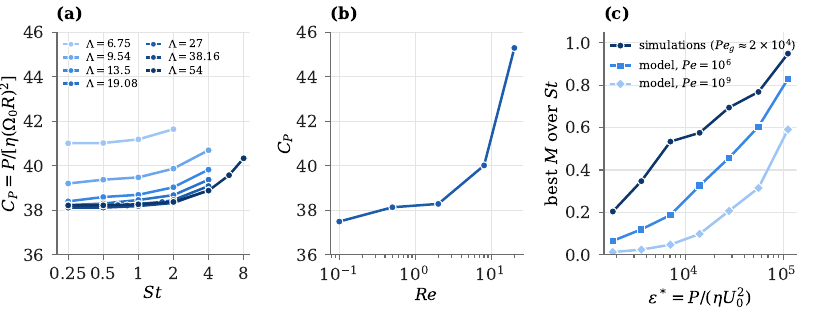}
\caption{Energy cost. (a) Power coefficient $C_P=P/[\eta(\Omega_0R)^2]$ versus the Strouhal number for the seven forcing ratios, $\Rey=2$; (b) $C_P$ versus the Reynolds number, $\St=1$, $\Lambda=27$; (c) best mixing efficiency over the sampled Strouhal numbers versus the dimensionless specific energy $\varepsilon^*$, for $\Lambda=6.75$, 9.54, 13.5, 19.08, 27, 38.16 and 54 (from left to right): simulations and lamellar model at $\Pe=10^6$ and $10^9$ (Sec.~\ref{sec:Pe}).}
\label{fig:energy}
\end{figure}

The rate of viscous dissipation per unit depth in the active zone $\mathcal A$ ($1.5\le x\le4.25$) is
\begin{equation}\label{eq:Pdef}
P(t)=\int_{\mathcal A}2\eta\,\mathsf D:\mathsf D\dd A,\qquad \mathsf D=\tfrac12\left[\nabla\bv+(\nabla\bv)^{\mathsf T}\right],
\end{equation}
where $\mathsf D:\mathsf D=D_{ij}D_{ij}$, and $P$ without argument denotes its average over a period. In the periodic regime, $P$, and with it the power injected by the arcs, follows from the superposition of Sec.~\ref{subsec:linear}. If $\bv\simeq U_0\bm u_b+\Omega(t)R\,\bm u_a$, the rate of strain is $\mathsf D=U_0\mathsf D_b+\Omega R\,\mathsf D_a$, and Eq.~\eqref{eq:Pdef} becomes
\begin{equation}
P(t)=\eta\left[a_{bb}U_0^2+2a_{ab}U_0\,\Omega(t)R+a_{aa}\,\Omega(t)^2R^2\right],\qquad a_{ij}=\int_{\mathcal A}2\,\mathsf D_i:\mathsf D_j\dd A,
\end{equation}
where the coefficients $a_{ij}$ depend only on the geometry. Over a period, $\Omega$ averages to zero and $\Omega^2$ to $\Omega_0^2/2$, so that
\begin{equation}
P=P_0+c_a\,\eta(\Omega_0R)^2,\qquad P_0=a_{bb}\,\eta U_0^2,\quad c_a=\tfrac12a_{aa},
\label{eq:power}
\end{equation}
independent of the period, and of the Reynolds number, as long as the superposition holds. The two coefficients are measured. A least-squares fit over the 37 runs at $\Rey=2$ and $\St\le4$ gives $P_0\simeq81\,\eta U_0^2$ (per unit depth) and $c_a=38.3$, with deviations below 4\,\%; their absolute values carry the grid uncertainty of about 10\,\% (Sec.~\ref{subsec:numerics}). The through-flow term $P_0$ is larger than that of a plane Poiseuille flow over the active zone, $12\,\eta U_0^2L_{a}/W\simeq33\,\eta U_0^2$ with $L_a=2.75$, because the arcs constrict the channel. At every forcing ratio, the power coefficient $C_P=P/[\eta(\Omega_0R)^2]$ varies by at most 4\,\% with the period over $\St=0.25$--4 (Fig.~\ref{fig:energy}a). Since $C_P=c_a+P_0/(\eta\Lambda^2U_0^2)$, it is slightly larger at $\Lambda=6.75$ (41), where the through-flow contributes relatively more. It departs from Eq.~\eqref{eq:power} where the quasi-steady superposition fails (Sec.~\ref{subsec:linear}): it increases by 3 and 5\,\% at $\St=6$ and 8 ($\Lambda=54$), beyond the window, and with inertia, from 37.5 at $\Rey=0.1$ to 38.3 at $\Rey=2$, 40.0 at $\Rey=8$ and 45.3 at $\Rey=20$ (Fig.~\ref{fig:energy}b).

The period therefore does not affect the energy cost. At a given forcing ratio the energy spent does not depend on $\St$, so the period should always be set at the optimum inside the window: forcing outside it costs the same energy for a much lower efficiency. In the efficiency map of Fig.~\ref{fig:emap}, the lines of constant power are horizontal, so that the map is directly an operating map. The forcing ratio, by contrast, sets the cost. The energy spent per unit volume of processed fluid, $\varepsilon=P/(U_0W)$, made dimensionless with $\eta U_0/W$, is $\varepsilon^*=C_P\Lambda^2\simeq P_0/(\eta U_0^2)+c_a\Lambda^2$; it ranges from $1.8\times10^3$ at $\Lambda=6.75$ to $1.1\times10^5$ at $\Lambda=54$. The best efficiency over the sampled periods increases with it (Fig.~\ref{fig:energy}c), from 0.20 to 0.35, 0.53, 0.58, 0.70, 0.77 and 0.95. The efficiency per unit energy, $M/\varepsilon^*$, however, decreases monotonically, by a factor of 13 over this range; having no optimum, it is not a useful criterion. The useful question is which forcing reaches a target efficiency at the lowest cost, at the P\'eclet number of the application. Interpolating in $\log\Lambda$, $M=0.5$ requires $\Lambda\approx13$ at the effective P\'eclet number of the simulations and, according to the lamellar model, $\Lambda\approx30$ at $\Pe=10^6$ and $\Lambda\approx48$ at $\Pe=10^9$, i.e.\ about five and fourteen times more energy. Given the errors of the model at $\Pe_g$ (up to 0.14, Sec.~\ref{sec:Pe}), these factors are estimates. A larger forcing ratio enlarges the saddle until it fills the mixing zone (Sec.~\ref{subsec:lcsLambda}), at a cost that grows as $\Lambda^2$. The period, which does not affect the cost, determines whether all the fluid passes close to the saddle.

\section{Discussion}\label{sec:discussion}
Taken together, the results show that mixing in the rotating arc-wall mixer is limited not by the strength of the chaos but by the fraction of the fluid that leaves the mixer before being stretched. Over the whole range explored, the chaotic saddle is strong and free of regular islands. What matters is therefore not whether the flow is chaotic, but how much of the fluid passes close enough to the saddle, and stays there long enough, to be stretched. Each control parameter acts on this fraction through a different coherent structure. The Strouhal number controls how the fluid is delivered to the saddle. If the forcing is too slow, the flow becomes a sequence of almost steady cellular flows whose cells let fluid through in pulses; if it is too fast, the fluid is held near the arcs without being folded. The window of Eq.~\eqref{eq:window} expresses these two failures in terms of two measured kinematic quantities, the shortest residence time and the amplitude of the cross-flow at mid-channel. The forcing ratio controls the size and strength of the saddle. With weak forcing, the saddle is a thin set near the arcs, and the fluid bypasses it through an unstirred passage and the near-wall regions; with strong forcing, it fills the mixing zone, which is why the optimum then moves to faster periods. The Reynolds number strengthens the saddle but also opens a wake behind the last arc. The P\'eclet number does not change the flow, but it sets the stretching needed for a striation to reach the diffusive scale, and therefore the contribution of the poorly stretched fluid to the outlet variance.

These results differ from the usual characterisation of open chaotic mixers by the escape rate and the fractal dimension of the saddle \cite{jung1993,lt2011,tel2005}. Here, the dimension of the unstable manifold is largest where the mixing collapses, with fast forcing, and the escape rate correlates with the efficiency with opposite signs in the Strouhal and forcing-ratio series: holding the fluid longer helps only if the fluid is stretched while it is held. The contrast of the FTLE field, a property of the velocity field over one time unit, correlates with the efficiency much better across all three series, because it directly measures the weakly stretched regions through which fluid escapes. The contrast is also lowest near the optimum at the reference forcing ratio. The optimum ($\St\approx0.7$) still differs from the maximum of the stretching rate ($\St=1$), at the grid-limited as at a resolved P\'eclet number. This difference comes from the pulsation of the outlet composition, i.e.\ from how the striations are arranged in time rather than from their thickness, and neither the saddle nor the lamellar model describes it.

The trends agree with the experiments in the same mixer \cite{kiwan2025}, although the geometry and the protocol differ in some details. The experiments found the optimum at $\St\approx0.7$ for a forcing ratio of 10. Interpolating $\tau_{\min}$ between the series at $\Lambda=9.54$ and 13.5, the window at $\Lambda=10$ is $0.30\lesssim\St\lesssim0.90$, which contains this optimum. The experiments also found that the efficiency increases monotonically with the forcing ratio, as found here for every sampled period; the shift of the optimum to faster forcing with increasing forcing ratio, predicted here by the window, was reported for a shear-thinning solution rather than for the Newtonian fluid. The separation they observed between the maxima of the stretching and of the mixing is discussed in Sec.~\ref{subsec:St}. The map of El Omari et al.\ \cite{PhysRevFluids.6.024502} has the same structure as Fig.~\ref{fig:emap}: a region of good mixing bounded, on the side of fast forcing, by a line through the origin and, on the side of weak forcing, by a minimum forcing ratio. Their minimum, $\Lambda\approx7.4$ for $\Pe=10^6$, is close to the critical forcing ratio $\Lambda_c\approx7$ at which the window closes, and the two are complementary. $\Lambda_c$ is kinematic and independent of $\Pe$, whereas their mixing condition accounts for the growth with the P\'eclet number of the forcing ratio needed for a useful efficiency, which the lamellar model confirms: with weak forcing much of the fluid is poorly stretched.

The comparisons in $\St$, $\Lambda$ and $\Rey$ are made at a resolution at which numerical diffusion acts like a molecular diffusivity with $\Pe_g\approx2\times10^4$. The window, the structure of the saddle and the independence of the cost from the period do not depend on this value, but the efficiency does. At the P\'eclet numbers of applications, only strong forcing mixes well, at an energy cost that grows as $\Lambda^2$. In this respect inertia is a secondary parameter: moderate inertia ($\Rey=8$) improves the efficiency by about 7\,\% for 4\,\% more power, whereas stronger inertia ($\Rey=20$) degrades it and increases the power by 18\,\%.

Several limitations should be kept in mind. The simulations are two-dimensional. In a mixer of finite span, the end walls add secondary flows and three-dimensional stretching, which should improve the mixing, shorten the residence time near the walls and possibly stir the regions that are weakly stretched here. The geometry and the protocol are fixed: the coefficient $\alpha$ and the shortest residence time, which set the window, are properties of this geometry, although the form of the two conditions is not. Both are kinematic: they are measured on the velocity field, $\alpha$ from the first harmonic of a single periodic run (Sec.~\ref{subsec:linear}) and $\tau_{\min}$ from tracers, without computing any concentration field. The threshold $S_c$, in contrast, is calibrated on the efficiency map, although any value between 0.7 and 1.2 classifies 30 of the 39 runs correctly (Sec.~\ref{subsec:Lambda}). For another geometry, the window therefore requires a few velocity computations and one calibration of $S_c$, not a new map. The map is sampled at seven forcing ratios and, except at $\Lambda=27$, at periods spaced by factors of two, and the Lagrangian structure is computed only along the three series through the reference point. The Lagrangian quantities are computed from the phase-averaged velocity and do not resolve the dynamics in the cells next to the walls. The scalar field is grid-limited, and the lamellar model used to extrapolate to larger P\'eclet numbers has systematic errors with fast forcing and at low Reynolds numbers. Finally, the efficiency is measured at a single station, downstream of the last arc.

\section{Conclusions}\label{sec:conclusions}
Using two-dimensional simulations, I have studied how a Newtonian fluid is transported and mixed in the rotating arc-wall mixer over wide ranges of the Strouhal number (0.25--8), the forcing ratio (6.75--54), the Reynolds number (0.1--20) and the P\'eclet number ($10^2$ up to the grid-limited regime). Throughout this range, the fluid passes through a chaotic saddle that contains no regular islands. The mixing is therefore not limited by a lack of chaos, but by the part of the fluid that crosses the mixer without being stretched much. Accordingly, the quantity that correlates best with the mixing efficiency is the contrast of the stretching field, i.e.\ how unevenly the stretching is distributed, rather than the stretching rate of the saddle, its escape rate or its fractal dimension.

Each control parameter acts on this poorly stretched fluid through a distinct coherent structure. Two kinematic conditions, based on the shortest residence time and on the cross-flow at mid-channel, bound a window of efficient periods, $W/(2U_0\tau_{\min})\lesssim\St\lesssim\alpha\Lambda/\pi$, with $\alpha\simeq0.28$. Below the window, the flow is a sequence of almost steady cellular flows whose cells let fluid through in pulses. Above it, the strokes are too short to fold the fluid, which is held near the arcs without being stretched. At the reference forcing ratio, the optimum, $\St\approx0.7$, is the slowest period that reverses every parcel at least once, and the one at which the mixed fluid is delivered most steadily. The window closes below $\Lambda_c\approx7$ and moves to faster forcing as the forcing ratio increases. At the same time, the saddle grows until it fills the mixing zone, so that with strong forcing the optimum moves to the upper limit of the window. Inertia strengthens the saddle but opens a weakly stirred wake behind the last arc. The efficiency is therefore insensitive to the Reynolds number in the creeping regime, is highest at moderate inertia ($\Rey\approx8$) and decreases beyond. The efficiency decreases only slowly with the P\'eclet number, as expected from exponential stretching. A lamellar model based on the measured stretching relates the grid-limited simulations ($\Pe_g\approx2\times10^4$) to applications, where the poorly stretched fluid accounts for an even larger part of the outlet variance.

As expected from the linearity of the creeping flow, the power injected by the arcs is $P\simeq P_0+38.3\,\eta(\Omega_0R)^2$, independent of the period. The forcing ratio therefore sets both the cost and the size of the saddle. It should be chosen from the target efficiency and the P\'eclet number of the material: according to the lamellar model, an efficiency of 0.5 requires about five times more energy at $\Pe=10^6$, and fourteen times more at $\Pe=10^9$, than in the simulations. The period does not affect the cost and should be set inside the window, where all the fluid passes close to the saddle. With a shear-thinning fluid, the speed of the walls matters in addition to their displacement \cite{ottino1}, because it sets the viscosity at the moving walls and with it the reach of the forcing \cite{burghelea_jnnfm}. Natural extensions of this work are three-dimensional flows, other arc geometries and forcing protocols, and simulations that resolve the P\'eclet number over the whole map.

\section*{Data availability}
The data that support the findings of this article (the concentration and velocity profiles at the measurement stations, the phase-averaged velocity fields and the Lagrangian diagnostics of every run) are openly available on Zenodo, \url{https://doi.org/10.5281/zenodo.24361990}.

\begin{acknowledgments}
This research used resources of the GLiCID Computing Facility (Ligerien Group for Intensive Distributed Computing, \url{https://doi.org/10.60487/glicid}, Pays de la Loire, France). This research did not receive any specific grant from funding agencies in the public, commercial, or not-for-profit sectors. The author used Claude (Anthropic), a large language model, to assist with running the simulations on the computing cluster, writing the analysis and plotting scripts, analysing the data, and drafting and editing the text. The author designed the study, checked the analyses and the results, revised the manuscript, and takes full responsibility for its content. The author declares no competing interests.
\end{acknowledgments}

\bibliography{refs}

\end{document}